\pdfoutput=1
\documentclass[aps,prd,amsmath,floats,floatfix, twocolumn,
superscriptaddress,nofootinbib,showpacs,longbibliography]{revtex4-1}
\usepackage[T1]{fontenc}
\usepackage[utf8]{inputenc}
\usepackage{txfonts}
\usepackage{verbatim}
\usepackage[dvipsnames, usenames]{xcolor}
\definecolor{linkcolor}{rgb}{0.0,0.3,0.5}
\usepackage[hypertexnames=false, unicode, colorlinks=true, linkcolor=linkcolor,
citecolor=linkcolor, filecolor=linkcolor,urlcolor=linkcolor,
pdfusetitle]{hyperref}
\usepackage[all]{hypcap}
\usepackage{graphicx}
\usepackage{xspace}
\usepackage{amssymb}
\usepackage{microtype}
\usepackage{subfigure}
\usepackage{appendix}

\usepackage{url}

\usepackage[english]{babel}
\usepackage{blindtext}

\usepackage[normalem]{ulem} 
\usepackage{bm} 
\usepackage{mathrsfs}
\usepackage{xspace} 

\DeclareMathAlphabet{\mathpzc}{OT1}{pzc}{m}{it}

\usepackage[nomessages]{fp}

\newcommand{\sk}[1]{}

\begin{document}
\title{Post-Newtonian theory-inspired framework for characterizing eccentricity in gravitational waveforms}
\newcommand{\KITP}{\affiliation{Kavli Institute for Theoretical Physics, University of California Santa Barbara, Kohn Hall, Lagoon Rd, Santa Barbara, CA 93106}} 
\newcommand{\TAPIR}{\affiliation{Theoretical AstroPhysics Including Relativity and Cosmology, California Institute of Technology, Pasadena, California, USA}}

\author{Tousif Islam}
\email{tousifislam@ucsb.edu}
\KITP
\TAPIR

\author{Tejaswi Venumadhav}
\affiliation{\mbox{Department of Physics, University of California at Santa Barbara, Santa Barbara, CA 93106, USA}}
\affiliation{\mbox{International Centre for Theoretical Sciences, Tata Institute of Fundamental Research, Bangalore 560089, India}}
 
\hypersetup{pdfauthor={Islam et al.}}

\date{\today}

\begin{abstract}
Characterizing eccentricity in gravitational waveforms in a consistent manner is crucial to facilitate parameter estimation, astrophysical population studies, as well as searches for these rare systems. 
We present a framework to characterize eccentricity directly from gravitational waveforms for non-precessing eccentric binary black hole (BBH) mergers using common modulations that eccentricity induces in all spherical harmonic modes of the signals. 
Our framework is in the spirit of existing methods that use frequency modulations in the waveforms, but we refine the approach by connecting it to state-of-the-art post-Newtonian calculations of the time evolution of the eccentricity. 
Using 39 numerical relativity (NR) simulations from the SXS and RIT catalogs, as well as waveforms obtained from the post-Newtonian approximation and effective-one-body (EOB) formalism, we show that our framework provides eccentricity estimates that connect smoothly into the relativistic regime (even up to $\sim 2M$ before merger). 
We also find that it is necessary to carry existing post-Newtonian calculations to an extra $0.5$PN order to adequately characterize existing NR simulations, and provide fits to the extra coefficient for existing simulations. 
We make the framework publicly available through the Python-based \texttt{gwModels} package. 
\end{abstract}

\maketitle

\section{Introduction}
Estimating eccentricity in binary compact object merger waveforms is a key challenge in gravitational wave (GW) astronomy~\cite{Tai:2014bfa}. The complexity arises from the absence of a unique definition of eccentricity in general relativity~\cite{Blanchet:2013haa}. As a result, various eccentricity definitions have been proposed, based on orbital quantities such as the radial separation between the compact objects or orbital frequencies~\cite{Mroue:2010re, Healy:2017zqj,Mora:2002gf}. Another set of eccentricity estimators relies on waveform features, such as amplitudes or instantaneous frequencies~\cite{Ramos-Buades:2021adz,Islam:2021mha,Ramos-Buades:2019uvh}. Different simulation or calculation frameworks, like numerical relativity (NR)~\cite{Mroue:2010re, Healy:2017zqj,Buonanno:2006ui,Husa:2007rh,Ramos-Buades:2018azo,Ramos-Buades:2019uvh,Purrer:2012wy,Bonino:2024xrv,Ramos-Buades:2022lgf} and point-particle perturbation theory (ppBHPT)~\cite{Warburton:2011fk,Osburn:2015duj,VanDeMeent:2018cgn,Chua:2020stf,Hughes:2021exa,Katz:2021yft,Lynch:2021ogr}, each adopt their own definitions, as do different waveform models~\cite{Tiwari:2019jtz, Huerta:2014eca, Moore:2016qxz, Damour:2004bz, Konigsdorffer:2006zt, Memmesheimer:2004cv,Hinder:2017sxy, Cho:2021oai,Chattaraj:2022tay,Hinderer:2017jcs,Cao:2017ndf,Chiaramello:2020ehz,Albanesi:2023bgi,Albanesi:2022xge,Riemenschneider:2021ppj,Chiaramello:2020ehz,Ramos-Buades:2021adz,Liu:2023ldr,Huerta:2016rwp,Huerta:2017kez,Joshi:2022ocr,Setyawati:2021gom,Wang:2023ueg,Islam:2021mha,Carullo:2023kvj,Nagar:2021gss,Tanay:2016zog,Paul:2024ujx,Manna:2024ycx} and remnant property estimators~\cite{Carullo:2023kvj}. Furthermore, astrophysical population models use their own eccentricity definitions~\cite{Vijaykumar:2024piy,Wen:2002km,Hamers:2021eir}. This lack of consistency across frameworks can lead to discrepancies in eccentricity measurements of eccentric signals, hindering the accurate interpretation of results and their comparison across different models.

For waveforms, the canonical post-Newtonian (PN) eccentricity parameter is often used to characterize eccentricity by fitting the waveform (e.g., obtained from NR simulations) to PN approximations~\cite{Moore:2016qxz,Gopakumar:1997bs,Damour:2004bz,Konigsdorffer:2006zt,Tessmer:2008tq,Arun:2009mc,Tanay:2016zog}. Currently, PN approximations for non-precessing eccentric binaries are known up to 3PN order~\cite{Arun:2009mc,Tanay:2016zog,Paul:2022xfy,Henry:2023tka,Sridhar:2024zms} and are sufficient for characterizing eccentricity in the early inspiral regime of the binary evolution. Furthermore, efforts have been made to standardize the definition of eccentricity across waveform and population models by utilizing waveform-based quantities that use amplitudes or instantaneous frequencies of the dominant quadrupolar mode at the apastron and periastron to construct estimators for eccentricity~\cite{2023PhRvD.108l4063R,Shaikh:2023ypz,Knee:2022hth,Boschini:2024scu}. While these approaches represent important progress, they have certain limitations. For example, PN approximations based on terms up to 3PN order break down in the late-inspiral regime, limiting the ability to robustly estimate eccentricity across the binary’s full evolution. Furthermore, 3PN approximations may not be sufficient to describe highly eccentric binaries. On the other hand, eccentricity estimators based on waveform quantities often do not recover the Newtonian limit~\cite{Ramos-Buades:2019uvh,Ramos-Buades:2021adz}. Additionally, they rely on spline or polynomial fits which are hard to extend closer to the merger~\cite{2023PhRvD.108l4063R,Shaikh:2023ypz,Knee:2022hth}.

We note that the primary purpose of defining eccentricity is to smoothly characterize its impact on gravitational radiation and remnant properties, thereby facilitating systematic model building and the inference of astrophysical properties from observed GW signals. For quasi-circular binaries, access to time-varying binary properties, such as spin evolution up to merger, enables the characterization of signals based on spin values at specific times or frequencies, which has assisted waveform modeling~\cite{Varma:2019csw, Yu:2023lml} and astrophysical inference~\cite{Mould:2021xst}. Similarly, obtaining robust estimates of eccentricity evolution up to merger is expected to yield comparable benefits for modeling and interpreting signals from eccentric binaries.

With that objective in mind, in this paper, we propose a simple yet robust framework to characterize eccentricity from waveform quantities up to merger by (i) building upon existing post-Newtonian calculations of the time evolution of eccentricity, and (ii) constructing an eccentricity estimator that leverages universal modulation features observed consistently across all waveform modes for eccentric non-precessing binaries in NR simulations~\cite{Islam:2024rhm,Islam:2024zqo,Islam:2024bza}. 
Thus our framework is both grounded in PN calculations, and informed by NR simulations. 
The modulations we use can be directly applied to quasicircular waveform models to efficiently generate eccentric models~\cite{Islam:2024zqo}, which makes the approach well-suited for developing waveform approximants. 

We provide the details of our framework in Section~\ref{sec:method}.
We demonstrate the effectiveness of the method using non-precessing eccentric waveforms from NR simulations in the SXS (\href{https://data.black-holes.org/waveforms/catalog.html}{https://data.black-holes.org/waveforms/catalog.html})~\cite{Boyle:2019kee,Hinder:2017sxy,LIGOScientific:2016ebw} and RIT (\href{https://ccrgpages.rit.edu/~RITCatalog/}{https://ccrgpages.rit.edu/~RITCatalog/})~\cite{Healy:2022wdn,Healy:2020vre} catalogs in in Section~\ref{sec:results}. In particular, we show that our eccentricity definition (i) recovers the Newtonian limit for small eccentricities, (ii) 
provides a smooth eccentricity time-evolution through PN-inspired fits, and (iii) yields meaningful eccentricity estimates even near merger. 
Finally, we discuss the implications of our results and their connections to existing work, limitations, and future improvements in Section~\ref{sec:implications}. In Section~\ref{sec:pnorders}, we show that our results offer insights into the missing higher-order terms in current PN calculations. 
In sections~\ref{sec:exi_to_egw} and ~\ref{sec:updated_egwfit}, we demonstrate that our framework connects to and enhances the robustness of other existing eccentricity estimators~\cite{2023PhRvD.108l4063R,Shaikh:2023ypz,Knee:2022hth}. 
Additionally, we show that the eccentricity evolution for non-spinning binaries can be effectively modeled by a simple overall power law (Sections~\ref{sec:ecc_evolve} and \ref{sec:gwEccEvNS}). Finally, we provide mappings between our proposed eccentricity definitions and those currently used in state-of-the-art waveform models (Section~\ref{sec:pn_eob}).
Our framework is available through the \texttt{gwModels} package at \href{https://github.com/tousifislam/gwModels}{https://github.com/tousifislam/gwModels}.

\section{Methods}
\label{sec:method}
\subsection{The universal modulation function}
The gravitational radiation (waveform) from a BBH merger is typically expressed as a complex time series constructed from two independent polarizations~\cite{Maggiore:2007ulw,Maggiore:2018sht}:
\begin{align}
h(t, \theta, \phi; \boldsymbol{\lambda}) &= h_{+}(t, \theta, \phi; \boldsymbol{\lambda}) - i h_{\times}(t, \theta, \phi; \boldsymbol{\lambda}).
\label{hmodes}
\end{align}
Here, $t$ represents time, while $\theta$ and $\phi$ are angles on the sky relative to the merger. The set of intrinsic parameters, $\boldsymbol{\lambda}$, includes quantities such as the masses and spins that describe the binary. For non-precessing eccentric BBHs, the set of parameters is given by $\boldsymbol{\lambda} := \{q, \chi_1, \chi_2, e_{\rm ref}, l_{\rm ref}\}$, where $q$ is the mass ratio, $\chi_1$ and $\chi_2$ are the dimensionless spin magnitudes of the larger and smaller black holes, respectively. Additionally, $e_{\rm ref}$ and $l_{\rm ref}$ refer to the eccentricity and the mean anomaly parameter.

The complex waveform is then decomposed into a superposition of $-2$ spin-weighted spherical harmonic modes with indices $(\ell, m)$:
\begin{align}
h(t, \theta, \phi; \boldsymbol{\lambda}) &= \sum_{\ell=2}^\infty \sum_{m=-\ell}^{\ell} h_{\ell m}(t; \boldsymbol\lambda) \, _{-2}Y_{\ell m}(\theta,\phi).
\label{hmodes}
\end{align}
Each spherical harmonic mode can be further decomposed into a real-valued amplitude $A_{\ell m}(t)$ and phase $\phi_{\ell m}(t)$, which yield an angular frequency $\omega_{\ell m}(t)$:
\begin{align}
    h_{\ell m}(t; q, e_{\rm ref}) & = A_{\ell m}(t) e^{i \phi_{\ell m}(t)}, \, {\rm and} \\
    \omega_{\ell m}(t; \boldsymbol{\lambda}) & = \frac{d\phi_{\ell m}(t)}{dt}.
\end{align}

\begin{table}[t]
\centering
\footnotesize 
\setlength{\tabcolsep}{10pt} 
\begin{tabular}{l c r}
 &  &  \\
\toprule
SXS:BBH:1355 & SXS:BBH:1356 & SXS:BBH:1357 \\
SXS:BBH:1358 & SXS:BBH:1359 & SXS:BBH:1360 \\
SXS:BBH:1361 & SXS:BBH:1362 & SXS:BBH:1363 \\
SXS:BBH:1364 & SXS:BBH:1365 & SXS:BBH:1366 \\
SXS:BBH:1367 & SXS:BBH:1368 & SXS:BBH:1369 \\
SXS:BBH:1370 & SXS:BBH:1371 & SXS:BBH:1372 \\
SXS:BBH:1373 & SXS:BBH:0108 & SXS:BBH:0319 \\
SXS:BBH:0320 & SXS:BBH:0321 & SXS:BBH:0322 \\
SXS:BBH:0323 & SXS:BBH:1149 & SXS:BBH:1169 \\
\toprule
\end{tabular}
\caption{SXS NR simulations~\cite{Hinder:2017sxy} used in this work.}
\label{tab:sxsdata_entries}
\end{table}

\begin{table}[t]
\centering
\footnotesize 
\setlength{\tabcolsep}{10pt} 
\begin{tabular}{l c r}
 &  &  \\
\toprule
RIT:eBBH:1282 & RIT:eBBH:1422 & RIT:eBBH:1330 \\
RIT:eBBH:1353 & RIT:eBBH:1376 & RIT:eBBH:1399 \\
RIT:eBBH:1445 & RIT:eBBH:1468 & RIT:eBBH:1491 \\
RIT:eBBH:1899 & RIT:eBBH:1763 & RIT:eBBH:1740 \\
\toprule
\end{tabular}
\caption{RIT NR simulations~\cite{Healy:2022wdn,Healy:2020vre} used in this work.}
\label{tab:ritdata_entries}
\end{table}

Ref.~\cite{Islam:2024rhm} empirically demonstrated that, for non-precessing eccentric BBHs, all spherical harmonic modes exhibit universal modulations caused by eccentricity. 
These modulations can be extracted from either the amplitudes $A_{\ell m}(t; \boldsymbol{\lambda})$ or the instantaneous frequency parameters $\omega_{\ell m}(t; \boldsymbol{\lambda})$ by comparing them to the values from the corresponding quasi-circular waveform characterized by $\boldsymbol{\lambda}^0 := \{q, \chi_1, \chi_2, e_{\rm ref}=0, l_{\rm ref}=0\}$:
\begin{align}
\xi_{\ell m}^{\omega}(t;\boldsymbol{\lambda}) & = b_{\ell m}^\omega \frac{\omega_{\ell m}(t; \boldsymbol{\lambda}) - \omega_{\ell m}(t; \boldsymbol{\lambda^0})}{\omega_{\ell m}(t; \boldsymbol{\lambda^0})}, \, {\rm and}
\label{eq:freq_mod} \\
\xi_{\ell m}^{A}(t;\boldsymbol{\lambda}) & = b_{\ell m}^A \frac{2}{\ell} 
\frac{A_{\ell m}(t; \boldsymbol{\lambda}) - A_{\ell m}(t; \boldsymbol{\lambda^0})}{A_{\ell m}(t; \boldsymbol{\lambda^0})}.
\label{eq:amp_mod}
\end{align}
where $b_{\ell m}^\omega\!$ and $b^{A}_{\ell m}\!$ are constants that can be chosen to enforce the desired limiting behavior -- in our case, to reduce to the Newtonian definition of eccentricity in the low eccentricity limit.
Ref.~\cite{Islam:2024rhm} also empirically shows that the amplitude and frequency modulations are related by: 
\begin{equation}
\xi_{\ell m}^{A}(t; \boldsymbol{\lambda}) = B \, \xi_{\ell m}^{\omega}(t; \boldsymbol{\lambda}),
\end{equation}
where the scaling factor $B \approx 0.9$. 
If we assume this consistency, we can define a common eccentric modulation function, denoted as $\xi(t; \boldsymbol{\lambda})$:
\begin{equation}
\xi(t; \boldsymbol{\lambda}) := \frac{\xi_{\ell m}^{A}(t; \boldsymbol{\lambda})}{B} = \, \xi_{\ell m}^{\omega}(t; \boldsymbol{\lambda}).
\label{eq:common_mod}
\end{equation}
We empirically find that $\xi_{22}^{A}(t; \boldsymbol{\lambda})$ provides the cleanest modulation function, as it is less affected by numerical errors in NR simulations, and hence we use it unless noted otherwise. 

\begin{figure*}
\includegraphics[width=\textwidth]{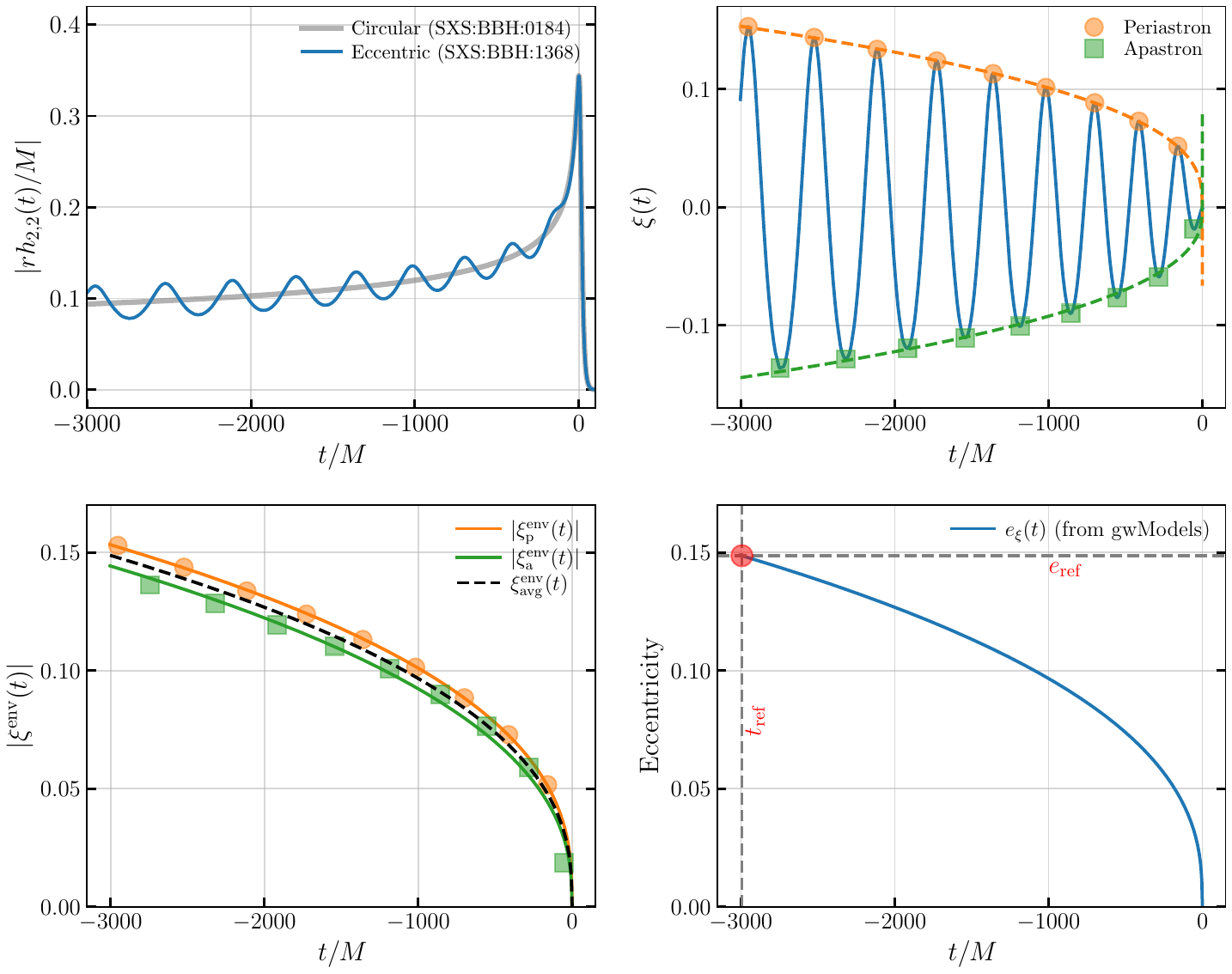}
\caption{We demonstrate the proposed procedure to estimate eccentricity $e_{\xi}(t)$ from an eccentric NR simulation. This framework is implemented in the \texttt{gwModels} package (\href{https://github.com/tousifislam/gwModels}{https://github.com/tousifislam/gwModels}). \textit{(Upper left panel)} We show the amplitudes of the dominant $(2,2)$ mode for the eccentric non-spinning BBH merger simulation \texttt{SXS:BBH:1368} (blue solid line) and the corresponding circular simulation \texttt{SXS:BBH:0184} (grey solid line). \textit{(Upper right panel)} We show the eccentric modulation function $\xi(t)$ (blue solid line) obtained from the amplitudes, as well as the maximas (orange circles) and minimas (green squares) corresponding to the periastrons and apastrons. \textit{(Lower left panel)} We show the envelopes $\xi^{\rm env}_{\rm p}(t)$ (orange solid line) and $\xi^{\rm env}_{\rm a}(t)$ (green solid line), obtained from analytical fits to the periastron and apastron values of $\xi(t)$ respectively, as well as the average envelope functions $\xi^{\rm env}_{\rm avg}(t)$ (black dashed line). \textit{(Lower right panel)} Finally, we present the eccentricity estimate $e_{\xi}(t)$ (blue solid line) from the average envelope. Details are in Section~\ref{sec:sxs1368}.
}
\label{fig:SXSBBH1368_example_ecc}
\end{figure*}

\subsection{Connection between the modulation function and eccentricity in the Newtonian limit}
In the Newtonian limit, eccentricity \( e_{\rm  N}(t) \), amplitude \( A_{22}(t) \), orbital phase \( \phi_{22}(t) \), and $(2,2)$ mode frequency \( \omega_{22}(t) \) are related as~\cite{Healy:2017zqj} (see also Appendix B of Ref.~\cite{Purrer:2012wy}):
\begin{equation}
\omega_{22}(t; \boldsymbol{\lambda}) = \omega_{22}(t; \boldsymbol{\lambda^0}) \left(1+\frac{3}{2} e_{\rm  N} \cos \phi_{22} \right)+\mathcal{O}\left(e_{\rm  N}^2\right),
\end{equation}
and
\begin{equation}
A_{22}(t; \boldsymbol{\lambda}) = A_{22}(t; \boldsymbol{\lambda^0})\left(1+\frac{3}{2} e_{\rm  N} \cos \phi_{22} \right)+\mathcal{O}\left(e_{\rm  N}^2\right).
\end{equation}
This can then be inverted to get a measure of eccentricity as ~\cite{Healy:2017zqj}:
\begin{equation}
e_{\omega}(t) := e_{\rm  N} \cos \phi_{22} = \frac{2}{3} \frac{\omega_{22}(t; \boldsymbol{\lambda}) - \omega_{22}(t; \boldsymbol{\lambda^0})}{\omega_{22}(t; \boldsymbol{\lambda^0})},
\label{eq:ecc_estimate}
\end{equation}
where the envelope of $e_{\omega}(t)$ gives $e_{\rm N}(t)$. This Newtonian eccentricity \( e_{\rm  N}(t) \) is uniquely defined using the radial separations between the two black holes. We immediately identify that $e_{\omega}(t)$ is simply $\xi_{\ell m}^{\omega}(t;\boldsymbol{\lambda})$ with
\begin{equation}
b^{\omega}_{\ell m} = 2/3. 
\end{equation}
A similar definition can be obtained from the amplitude as well with $b^{A}_{\ell m} = 2/3$. 
In other words, the common eccentricity modulation function $\xi(t; \boldsymbol{\lambda})$ (with $b^{A}_{\ell m} = 2/3$) reduces to the eccentricity measure in the Newtonian limit. 
Analogous results for other modes can also be derived by following the approaches outlined here, starting from the Newtonian results presented in Refs.~\cite{Healy:2017zqj,Maggiore:2007ulw,Maggiore:2018sht}.
Motivated by this, we now develop a framework to characterize eccentricity based on the calculated \( \xi(t; \boldsymbol{\lambda}) \) time series. We elaborate on our framework in the following sections.

\subsection{Detailed framework for characterizing eccentricity from modulation function}
We first note that the modulation time-series $\xi(t; \boldsymbol{\lambda})$ is a real-valued function that exhibits a sinusoidal decay behavior. 
It is possible to identify the envelope of this decaying time-series. 
In fact, $\xi(t; \boldsymbol{\lambda})$ has two envelopes: the upper envelope $\xi^{\rm env}_{\rm p}(t)$ corresponds to the values associated with periastron, and the lower envelope $\xi^{\rm env}_{\rm a}(t)$ corresponds to the values associated with apastron. 
The absolute values of the envelopes, i.e., $\xi^{\rm env}_{\rm p}(t)$ and $-\xi^{\rm env}_{\rm a}(t)$ are monotonically decreasing, positive-valued functions of time. Therefore, they provide an effective means of characterizing eccentricity evolution.

Within the PN approximation, the time evolution of the canonical eccentricity parameter (for non-spinning BBH mergers) is given by
\begin{equation}
e_{\rm 3PN}(\tau, \tau_0, \nu, e_{0,\rm 3PN}) = e_{0,\rm 3PN} \times g_{\rm 3PN}(\tau, \tau_0, \nu),
\label{eq:3PN_ecc}
\end{equation}
where $e_{0, \rm 3PN}$ is the initial PN eccentricity, $\nu \equiv q/(1+q)^2$ is the symmetric mass ratio, and $\tau \equiv \nu (t_c - t)$ is the PN characteristic time with the characteristic reference time $\tau_0 \equiv \nu (t_c - t_{\rm ref})$ (where $t_{\rm ref}$ is the reference time and $t_c$ is a PN parameter that is often chosen to represent the time of merger.). Here, $g_{\rm 3PN}(\tau, \tau_0, \nu)$ is the eccentricity evolution function. The expression of $g_{\rm 3PN}(\tau, \tau_0, \nu)$ up to 3PN order looks like~\cite{Moore:2016qxz}:
\begin{equation}
\begin{aligned}
g_{\rm 3PN}(\tau)=\left(\frac{\tau}{\tau_0}\right)^{19 / 48}\left\{1+\left(-\frac{4445}{6912}+\frac{185}{576} \nu\right)\left(\tau^{-1 / 4}-\tau_0^{-1 / 4}\right) + .. \right\}.
\label{eq:fit_func}
\end{aligned}   
\end{equation}
We provide the full expression in Appendix~\ref{sec:3pn}. In the Newtonian limit, Eq.(\ref{eq:3PN_ecc}) reduces to the following one:
\begin{equation}
e_{\rm Newt}(\tau, \tau_0, \nu, e_{0,\rm Newt}) = e_{0,\rm Newt} \times \left(\frac{\tau}{\tau_0}\right)^{19 / 48},
\label{eq:Newt_ecc}
\end{equation}
where $e_{0, \rm Newt}$ is the initial eccentricity parameter.

To obtain a continuous representation of the envelopes $\xi^{\rm env}_{\rm p}(t)$ and $\xi^{\rm env}_{\rm a}(t)$, we first identify the discrete maxima $\xi^{\rm env}_{\rm p}(t_{\rm p})$ and minima $\xi^{\rm env}_{\rm a}(t_{\rm a})$ in the eccentric modulation functions, where $t_{\rm p}$ and $t_{\rm a}$ are the times corresponding to periastron and apastron, respectively. 
Our intuition is that the envelopes are linked to eccentricity, and hence should decay with time in a manner that's similar to Eq.~(\ref{eq:3PN_ecc}).
Hence we fit both the numerically obtained values of $\xi^{\rm env}_{\rm p}(t_{\rm p})$ and $\xi^{\rm env}_{\rm a}(t_{\rm a})$ to the form in Eq.~(\ref{eq:3PN_ecc}), replacing the prefactor $e_{0,\rm 3PN}$ with $\xi^{\rm env}_{{\rm p/a, }0,\rm 3PN}$ (where $\xi^{\rm env}_{{\rm p/a, }0,\rm 3PN} \propto e_{0,\rm 3PN}$) with {\rm p/a} indicating periastron and apastron respectively. This fit will then involve only one free parameter, $\xi^{\rm env}_{{\rm p/a, }0,\rm 3PN}$, for each envelope. 

When we carry this through, we find that 3PN expressions are not sufficient to capture the time evolution close to the merger for highly eccentric binaries (see Figure \ref{fig:SXSBBH1371_diff_order_PNfit} later). 
Therefore, we incorporate an additional term in our fitting function and use the following ansatz for the eccentricity evolution (we discuss the motivation behind this choice in detail in Section~\ref{sec:pnorders}), with $\xi^{\rm env}_0$ representing the initial value of the envelop:
\begin{equation}
\xi^{\rm env}_{\rm fit}(\tau, \tau_0, \nu, \xi_0) = \xi^{\rm env}_0 \times f(\tau, \tau_0, \nu),
\label{eq:fit_func1}
\end{equation}
where
\begin{equation}
f(\tau, \tau_0, \nu) = g_{\rm 3PN}(\tau, \tau_0, \nu) + \left( \frac{\tau}{\tau_0} \right)^{19 / 48} \left( c_{-1/8} \tau_0^{-1 / 8} \right).
\label{eq:fit_func2}
\end{equation}
For a given BBH merger, $\nu$ is known and $t_{\rm ref}$ (and therefore $\tau_0$) is already chosen. Thus, the only fit parameters are $\xi^{\rm env}_0$ and $c_{-1/8}$. For better fits, we allow the value of $c_{-1/8}$ to be different between the two envelopes.

At this point, we have fitted expressions for both the envelopes, and we could develop an eccentricity estimator using either of them, but we choose to use the average envelope to reflect contributions from both the apastron and periastron points:
\begin{equation}
\xi^{\rm env}_{\rm avg}(t) = \frac{\xi^{\rm env}_{\rm p}(t) + \xi^{\rm env}_{\rm a}(t)}{2},
\end{equation}
and this becomes our fiducial eccentricity parameter:
\begin{equation}
e_{\xi}^{\rm avg}(t) :=  \xi^{\rm env}_{\rm avg}(t).
\label{eq:egw}
\end{equation}
The reference eccentricity at the reference time $t_{\rm ref}$ then becomes: $e_{\rm ref}^{\rm avg} = e_{\xi}^{\rm avg}(t = t_{\rm ref})$.

In Appendix~\ref{sec:qk}, using the quasi-Keplerian description of the eccentric binaries in PN framework, we demonstrate that $e_{\xi}(t)$ correctly recovers the Newtonian limit in the low-eccentricity regime.

\subsection{Characterizing eccentricity using other methods}
\label{sec:ecc_other_methods}
One of the most popular definitions of eccentricity in the literature is based on the instantaneous frequency of the dominant $(2,2)$ mode, $\omega_{22}(t)$. It is expressed as~\cite{Mora:2002gf,Ramos-Buades:2022lgf}: 
\begin{equation} 
e_{\omega_{22}}(t) = \frac{\sqrt{\omega_{22}^{\mathrm{p}}(t)} - \sqrt{\omega_{22}^{\mathrm{a}}(t)}}{\sqrt{\omega_{22}^{\mathrm{p}}(t)} + \sqrt{\omega_{22}^{\mathrm{a}}(t)}}, 
\label{eq:ecc_omega_22}
\end{equation} 
where $\omega_{22}^{\mathrm{p}}(t)$ and $\omega_{22}^{\mathrm{a}}(t)$ correspond to the frequencies at the periastron and apastron, respectively. However, as pointed out in Ref.~\cite{Ramos-Buades:2022lgf}, this definition does not correctly reduce to the Newtonian limit. To address this, a modified eccentricity parameter is introduced as~\cite{Ramos-Buades:2022lgf}:
\begin{equation} 
e_{\mathrm{gw}} = \cos (\Psi / 3) - \sqrt{3} \sin (\Psi / 3), 
\label{eq:ecc_gw}
\end{equation}
with
\begin{equation} 
\Psi = \arctan \left( \frac{1 - e_{\omega_{22}}^2}{2 e_{\omega_{22}}} \right). 
\end{equation}
Note that, in practice, $\omega_{22}^{\mathrm{p}}(t)$ and $\omega_{22}^{\mathrm{a}}(t)$ are discrete values in time, corresponding to specific instances where periastron and apastron occur. Therefore, one first needs to identify these discrete $\omega_{22}^{\mathrm{p}}(t)$ and $\omega_{22}^{\mathrm{a}}(t)$ values. After that, either a fit or an interpolant must be constructed to track the behavior of these parameters across time smoothly. Similar definitions based on the periastron and apastron have also appeared in Ref.~\cite{Boschini:2024scu}.

\begin{figure}
\includegraphics[width=\columnwidth]{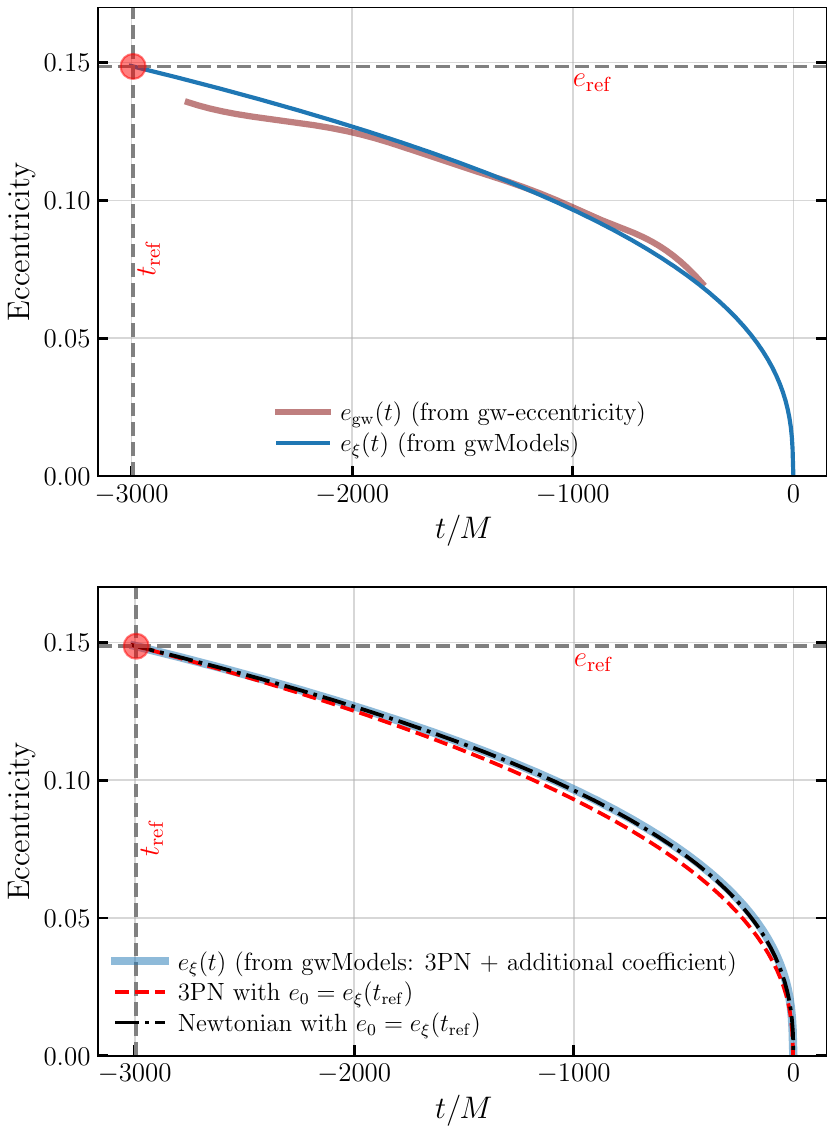}
\caption{(\textit{Upper panel}) We show the eccentricities estimated for the NR simulation \texttt{SXS:BBH:1368} using different methods implemented in \texttt{gwModels} (\href{https://github.com/tousifislam/gwModels}{https://github.com/tousifislam/gwModels}) and \texttt{gw-eccentricity} packages (with default spline fits). (\textit{Lower panel}) We show the eccentricity evolution obtained from the modulation function in Figure~\ref{fig:SXSBBH1368_example_ecc} as well as the expected eccentricity evolution in Newtonian and 3PN order. Details are in Section~\ref{sec:sxs1368}.}
\label{fig:SXSBBH1368_ecc_sanity_check}
\end{figure}

\subsection{Code availability}
We have implemented our framework of obtaining eccentricity directly from waveforms in a publicly available Python package named \texttt{gwModels}. The package is accessible at \href{https://github.com/tousifislam/gwModels}{https://github.com/tousifislam/gwModels}. This method can be used by calling the \texttt{EstimateEccentricity} class. A simple demonstration is provide in Appendix~\ref{sec:code}.

\section{Demonstrating the framework on NR data}
\label{sec:results}
We now demonstrate our eccentricity estimation framework using non-precessing eccentric NR waveforms obtained from SXS and RIT catalogs. In Tables~\ref{tab:sxsdata_entries} and ~\ref{tab:ritdata_entries}, we list these simulations. Additionally, we will compare our results with other definitions of eccentricity.

\subsection{Demonstration of the framework using \texttt{SXS:BBH:1368}}
\label{sec:sxs1368}
We first apply our method on a eccentric non-spinning BBH merger simulation \texttt{SXS:BBH:1368} (Figure~\ref{fig:SXSBBH1368_example_ecc}). This simulation is characterized by mass ratio $q=2$, spin on the primary $\chi_1=0.0$, spin on the secondary $\chi_2=0.0$, and eccentricity $e_{\rm ref}=0.21$ (as obtained from the simulation metadata provided at \href{https://data.black-holes.org/waveforms/catalog.html}{https://data.black-holes.org/waveforms/catalog.html}). The corresponding quasi-circular BBH merger simulation is \texttt{SXS:BBH:0184}. We align the quadrupolar mode of both waveforms at the merger time $t=0$, denoted by the maximum amplitude (Figure~\ref{fig:SXSBBH1368_example_ecc}, upper left panel).

\begin{figure}
\includegraphics[width=\columnwidth]{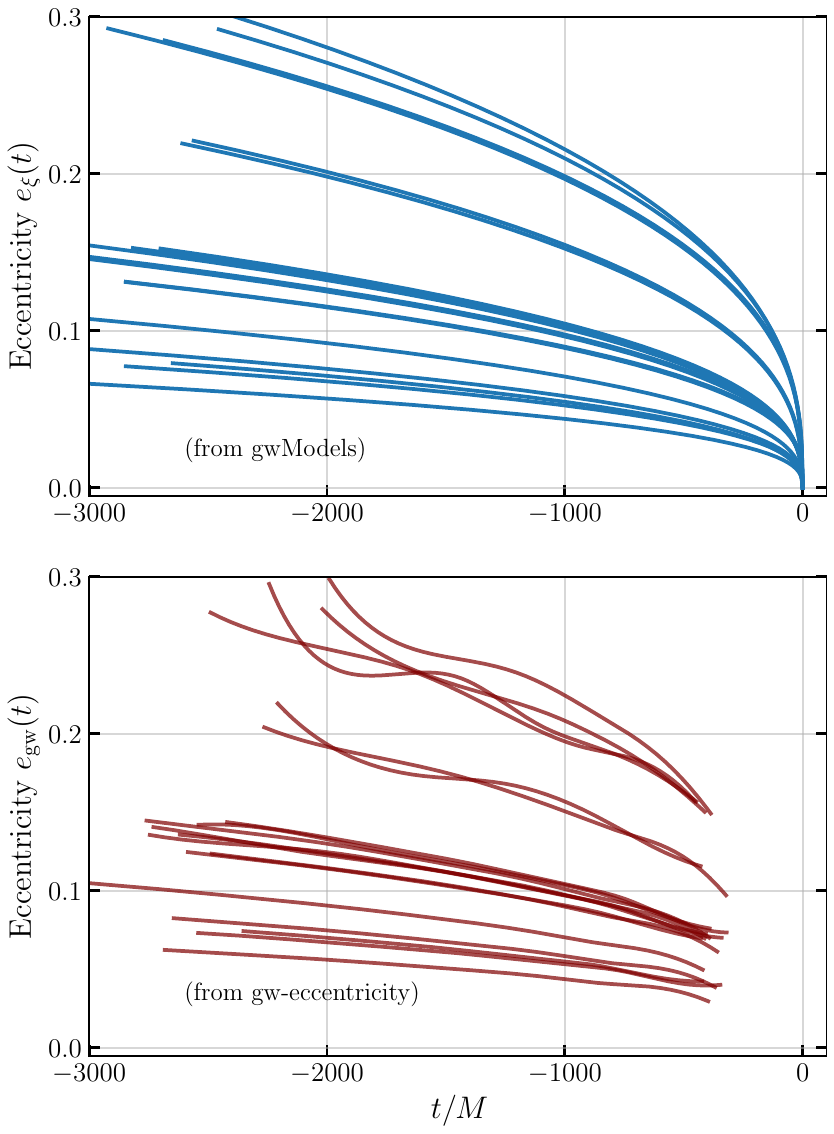}
\caption{We show the eccentricity estimated for a total of 19 SXS non-spinning eccentric BBH simulations with mass ratios $1 \leq q \leq 3$ from Ref.~\cite{Hinder:2017sxy} using both the \texttt{gwModels} (upper panel; blue lines; \href{https://github.com/tousifislam/gwModels}{https://github.com/tousifislam/gwModels}) and \texttt{gw-eccentricity} (lower panel; maroon lines) implementations. 
While \texttt{gw-eccentricity} uses amplitude/frequencies to estimate eccentricities, \texttt{gwModels} takes the universal eccentric modulation function as its input. Details are in Section~\ref{sec:sxsnospin}.}
\label{fig:gwModels_vs_gweccentricity}
\end{figure}

We then use Eq.~(\ref{eq:amp_mod}) and Eq.(\ref{eq:common_mod}) to compute the eccentric modulation function $\xi(t; \boldsymbol{\lambda})$ (Figure~\ref{fig:SXSBBH1368_example_ecc}, upper right panel). 
Next, we identify the points corresponding to the periastron and apastron (using the peak finder module \texttt{scipy.signal.find\_peaks}). 
These points correspond to the local maximas and minimas in the $\xi(t; \boldsymbol{\lambda})$ time-series. 
We fit the absolute values of the maximas and minimas using Eq.~(\ref{eq:fit_func1}), which provides the upper envelope $\xi^{\rm env}_{\rm p}(t)$ and lower envelope $\xi^{\rm env}_{\rm a}(t)$. 
The average envelope $\xi^{\rm env}_{\rm avg}(t)$ is then calculated, and all these envelopes are shown in the lower left panel of Figure~\ref{fig:SXSBBH1368_example_ecc}. 
Finally, we use Eq.~(\ref{eq:egw}) to compute the time-dependent eccentricity $e_{\xi}(t)$, whose functional form always decreases monotonically.
We find that the reference eccentricity $e_{\xi, \rm ref}=0.133$ which is almost 1.5 times smaller than the eccentricity value quoted in the simulation metadata. Note that, metadata eccentricity is obtained by fitting early orbital quantities to a chosen PN expression and those approximations may be responsible for the difference. We discuss this point more in next Sections.

At this point, it is worth mentioning a small issue with the approach. 
We find that it is much easier to fit the upper envelope than the lower one; for high eccentricities, we observe small differences between the numerical values and the best-fit lower envelopes around merger (also seen in the lower left panel of Figure~\ref{fig:SXSBBH1368_example_ecc}). 
This issue arises because, as eccentricity increases, the minima corresponding to the apastron become flatter, which makes it harder to identify the correct peak. 
In such cases, one may opt to use only the upper envelope to compute the eccentricity (although we have not adopted this approach). Despite the challenges with the lower envelope for high eccentricities, the fits provide a good representation of the modulation for the majority of simulations.

\begin{figure}
\includegraphics[width=\columnwidth]{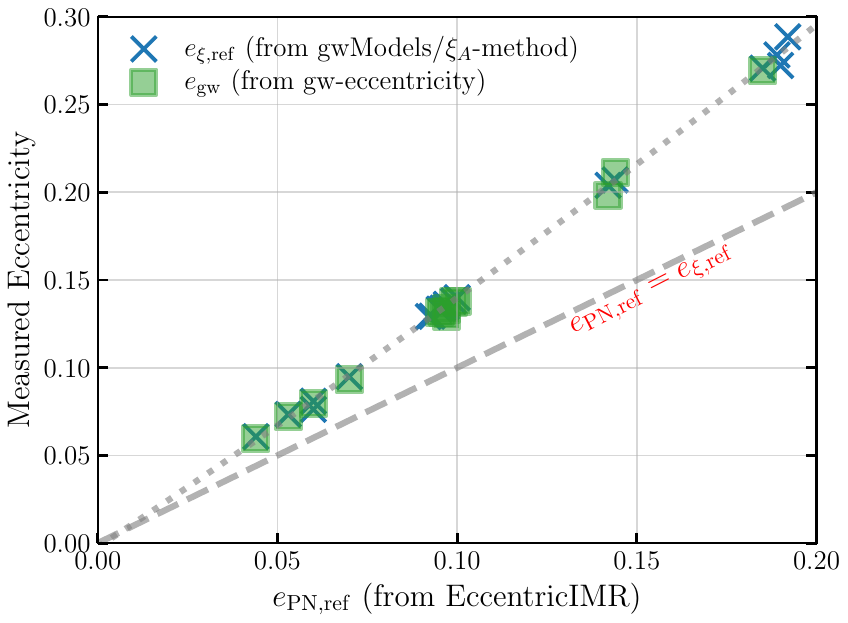}
\caption{We show the relationship between the estimated PN reference eccentricities for the 19 non-spinning SXS NR data (as reported in Ref.~\cite{Hinder:2017sxy}), our eccentricity estimates $e_{\xi, \rm ref}$ as well as $e_{\rm gw}$ (defined in Ref.~\cite{Ramos-Buades:2022lgf}) for the same data. Details are in Section~\ref{sec:sxsnospin}.}
\label{fig:pnecc_vs_eccxi}
\end{figure}

Next, we compare our eccentricity estimates \( e_{\xi}(t) \) (solid blue line) for the NR simulation \texttt{SXS:BBH:1368} using other definitions (\( e_{\omega_{22}} \) and \( e_{\rm gw} \)) discussed in Section~\ref{sec:ecc_other_methods} (Figure~\ref{fig:SXSBBH1368_ecc_sanity_check}; upper panel). To compute \( e_{\omega_{22}} \) (solid cyan line) and \( e_{\rm gw} \) (solid maroon line), we use the \texttt{gw-eccentricity} package with default spline fits~\cite{Shaikh:2023ypz}, which relies on orbital frequency and spline fits. We find that \texttt{gw-eccentricity} provides eccentricity estimates only up to \( t = -500M \), as the spline fits become unreliable afterward. Conversely, our framework can provide meaningful eccentricity estimates even at \(\sim 1M \) before the merger. The last non-zero eccentricity is obtained at \( t=-1.02M \), with an estimated value of \( e_{\xi}(t = -1.02) = 0.00028 \). Additionally, \texttt{gw-eccentricity} shows unphysical oscillations early in the waveform and around \( t = -1000M \). Otherwise, at intermediate times, \( e_{\rm gw}(t) \) values are close to \( e_{\xi}(t) \).

We then compare the eccentricity evolution \( e_{\xi}(t) \) with both the expected Newtonian, using Eq.(\ref{eq:Newt_ecc}), and 3PN evolution, using Eq.(\ref{eq:3PN_ecc}) (Figure~\ref{fig:SXSBBH1368_ecc_sanity_check}; lower panel). The initial eccentricity in the Newtonian and 3PN evolution is set to match the initial eccentricity in the \( e_{\xi}(t) \) time-series. We find that the eccentricity evolution, for this particular scenario, follows the Newtonian evolution more closely than the 3PN evolution. This difference may be due to the summation being truncated at an inconvenient PN order, as noted in several studies (e.g. Ref.~\cite{Wang:2024jro}). 
While we have not performed this experiment in this paper, it is possible that such issues could be mitigated by performing a resummation of the 3PN eccentricity evolution expression using a suitable Padé approximant~\cite{Bender1999}.

\begin{figure}
\includegraphics[width=\columnwidth]{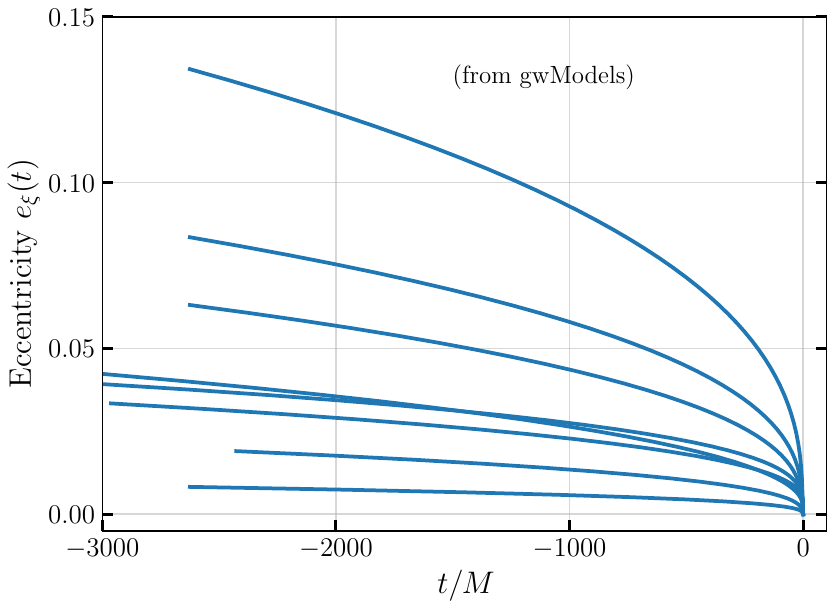}
\caption{We show the eccentricity estimated for a total of 8 SXS (anti)aligned-spin eccentric BBH simulations with mass ratios $1.22 \leq q \leq 5$ using the \texttt{gwModels} (\href{https://github.com/tousifislam/gwModels}{https://github.com/tousifislam/gwModels}) implementation. Details are in Section~\ref{sec:sxsspin}.}
\label{fig:gwModels_AlignedSpin_eccentricity}
\end{figure}

\subsection{Non-spinning SXS simulations}
\label{sec:sxsnospin}
Next, we apply our framework on 19 publicly available non-spinning eccentric BBH merger simulations from the SXS catalog (see Table~\ref{tab:sxsdata_entries}). 
These simulations have mass ratios $1 \leq q \leq 3$ from Ref.~\cite{Hinder:2017sxy}. 
We calculate the eccentricity time-series for these simulations using both the \texttt{gwModels} and \texttt{gw-eccentricity} implementations. 
We show these results in Figure~\ref{fig:gwModels_vs_gweccentricity}. 
As in the demonstration example, the \texttt{gwModels} implementation yields an eccentricity parameter that varies smoothly with time by construction.
In multiple cases, the eccentricity estimates that \texttt{gw-eccentricity} yields show oscillations that appear unphysical.
In all these cases, our eccentricity estimates \( e_{\xi}(t) \) yield meaningful values until very close to the merger—typically until \( 1M \) to \( 2M \) before the merger, with the last eccentricity measurements ranging from \( \sim 10^{-5} \) to \( 10^{-4} \).

For these 19 non-spinning eccentric BBH merger simulations, we also use the reference PN eccentricities from Ref.~\cite{Hinder:2017sxy}. These PN eccentricities are computed by matching the PN waveform frequency (derived from the \texttt{EccentricIMR} model~\cite{Hinder:2017sxy}) to NR over a single radial period, centered around a reference time corresponding to a dimensionless frequency of \( x = x_{\rm ref} \), where \( x = \omega_{\rm orb}^{2/3} \). Here, \( \omega_{\rm orb} \) represents the orbital frequency and is extracted from the (2,2) mode frequency as \( \omega_{\rm orb} = \omega_{22}/2 \). Ref.~\cite{Hinder:2017sxy} employed \( x_{\rm ref} = 0.075 \), chosen as the lowest frequency common to all the waveforms, which, for most simulations, corresponds to a time near the start of the simulation. \texttt{EccentricIMR} model uses 3PN conservative dynamics and 2PN adiabatic radiation reaction. While this approach leads to suboptimal long-term evolution and notable deviations from NR over time, the model matches NR well within a single orbit~\cite{Hinder:2008kv,Hinder:2017sxy}. Therefore, Ref.~\cite{Hinder:2017sxy} matched the PN waveform to NR over a single radial orbit. 

To compare our eccentricity estimates with these values, for each simulation, we first calculate the time corresponding to the dimensionless frequency of \( x = x_{\rm ref} = 0.075 \). Since the orbital frequency (and thus \( x \)) oscillates in eccentric simulations, we utilize the average dimensionless frequency. We then evaluate our eccentricity fits (as well as the eccentricity fits from \texttt{gw-eccentricity}) at that specific time and obtain the corresponding reference eccentricity from our method. In Figure~\ref{fig:pnecc_vs_eccxi}, we show that our reference eccentricity estimates (as well as the eccentricity estimates from \texttt{gw-eccentricity}) for these simulations are larger than the PN estimates reported in Ref.~\cite{Hinder:2017sxy}. 

\begin{figure}
\includegraphics[width=\columnwidth]{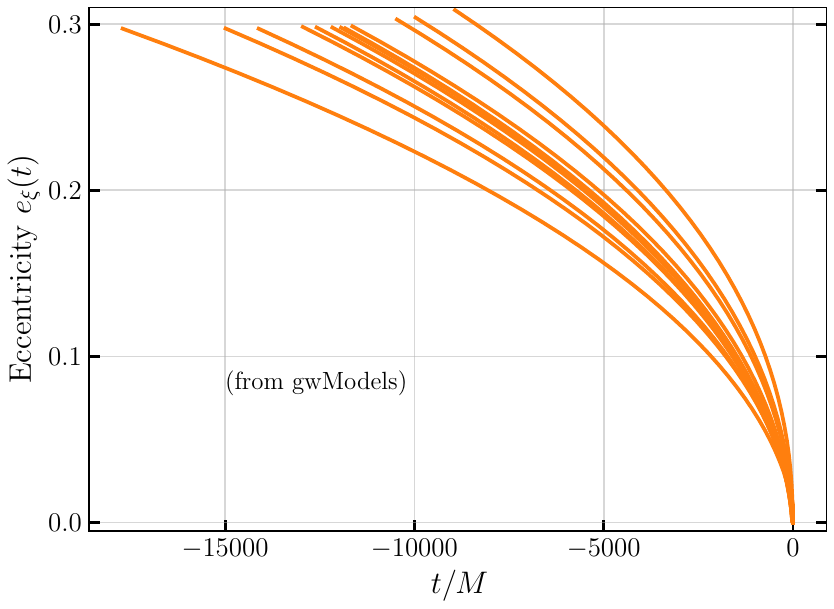}
\caption{We show the eccentricity estimated for a total of 12 publicly available long RIT eccentric BBH simulations with mass ratios $1 \leq q \leq 4$ using the \texttt{gwModels} (\href{https://github.com/tousifislam/gwModels}{https://github.com/tousifislam/gwModels}) implementation. Details are in Section~\ref{sec:rit}.}
\label{fig:gwModels_longRIT_eccentricity}
\end{figure}

However, it is important to note that we should not expect an exact match, as our estimates incorporate full 3PN terms in the non-spinning limit and an additional higher-order PN term, while the \texttt{EccentricIMR} model employs only 3PN conservative dynamics and 2PN adiabatic evolution for \( x(t) \) and \( e(t) \). Additionally, both $e_{\xi}$ and $e_{\rm gw}$ are designed to recover the Newtonian limit. However, this does not guarantee agreement with the PN eccentricity across all regimes. 
The following empirical fit approximately captures the relation between the estimators:
\sk{We quantify the departure through following analytical fit:}
\begin{equation}
    e_{\xi,\rm ref} \approx e_{\rm gw,\rm ref} \approx -0.002 + 1.21 \times e_{\rm PN,ref} + 0.6 \times e_{\rm PN,ref}^2.
\end{equation}

\begin{figure}
\includegraphics[width=\columnwidth]{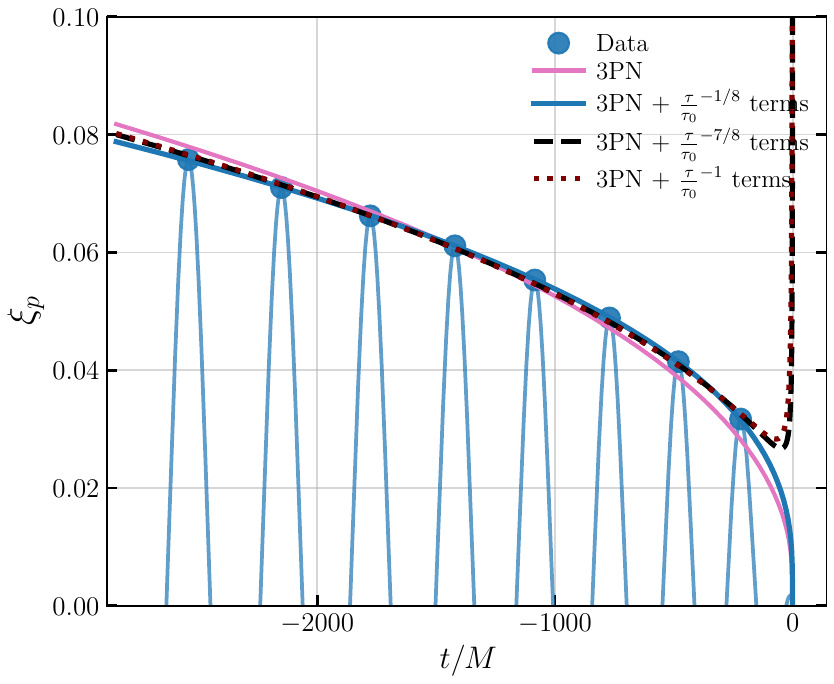}
\caption{We show the different fits, corresponding to various additional terms on top of the known 3PN expression given in Eq.~\ref{eq:3PN_ecc}, to the upper envelope $\xi_{\rm p}^{\rm env}$ of the modulation function for \texttt{SXS:BBH:1371}. We only show the upper half of the modulation function $\xi(t)$ for visual clarity. Details are in Section~\ref{sec:pnorders}.}
\label{fig:SXSBBH1371_diff_order_PNfit}
\end{figure}

\subsection{Spinning SXS simulations}
\label{sec:sxsspin}
Next, we apply our framework to 8 eccentric NR simulations with aligned and anti-aligned spins from the SXS catalog (see Table~\ref{tab:sxsdata_entries}), covering mass ratios from $q=1.22$ (similar to GW150914~\cite{LIGOScientific:2016ebw}) to $q=5$. Despite not incorporating spin effects in our eccentricity evolution model, we find that it still provides efficient eccentricity estimates. This is because our analytical fitting function can still fit the periastron and apastron envelops pretty well. However, in the future, we plan to include spin effects, particularly in the $g_{\rm PN}(\tau, \tau_0, \nu, e_0)$ expression using results from Refs.~\cite{Paul:2022xfy,Henry:2023tka}. Figure~\ref{fig:gwModels_AlignedSpin_eccentricity} shows the eccentricity evolution $e_{\xi}(t)$ for these binaries. As before, we find smooth monotonically decreasing eccentricity evolution. 

\subsection{RIT simulations}
\label{sec:rit}
We further characterize the eccentricity of 12 long RIT non-spinning and aligned/anti-aligned spin eccentric NR simulations (see Table~\ref{tab:ritdata_entries}), with mass ratios ranging from $q=1$ to $q=4$ for the non-spinning cases, and a fixed mass ratio of $q=1$ for the spinning cases. These simulations are $\sim 8000M$ to $\sim 18000M$ long in duration and the quoted eccentricities in the simulation metadata varies from $0.11$ to $0.19$. Our framework demonstrates robust performance even in these scenarios (Figure~\ref{fig:gwModels_longRIT_eccentricity}). We find that the initial estimated eccentricities (using \texttt{gwModels}) of these simulations are mostly around $0.2$. Additionally, we apply our framework to a set of five highly eccentric (anti)-aligned spin simulations with moderate duration (mostly $\sim 3000M$; not shown in the Figure). One such simulation is \texttt{RIT:eBBH:1828}, characterized by $q=1$, $\chi_1=0.8$, and $\chi_2=0.0$. The quoted initial eccentricity in the simulation metadata is $0.4375$, while our estimation yields $0.4875$. We recover similarly large initial eccentricities for the other four simulations (\texttt{RIT:eBBH:1786}, \texttt{RIT:eBBH:1807}, \texttt{RIT:eBBH:1883} and \texttt{RIT:eBBH:1862}) as well. This suggests that our framework can efficiently characterize moderately high eccentric waveforms too.

\section{Implications}
\label{sec:implications}
We now outline the implications of our work within the PN framework and in estimating alternative eccentricity parameters, such as $e_{\omega_{22}}$ and $e_{\rm gw}$. Using previous eccentricity estimates from NR data (Section~\ref{sec:results}), we also present an approximate model for eccentricity evolution in non-spinning binaries. 

\subsection{Insights into higher order PN terms}
\label{sec:pnorders}
Our eccentricity measurement framework, outlined in Section~\ref{sec:method}, is rooted in the 3PN expression for the canonical eccentricity parameter $e_t$ (check Eq.(\ref{eq:3PN_ecc}) of Ref.~\cite{Moore:2016qxz}). However, through numerical experiments, we observe that the 3PN expression is insufficient to capture the envelope of the modulation functions (i.e. $\xi_{\rm p}^{\rm env}$ and $\xi_{\rm a}^{\rm env}$). As the mass ratio and eccentricity increase, the fits worsen. Therefore, we decide to incorporate higher-order terms into our fitting ansatz (see Eq.~\ref{eq:fit_func2}).

\begin{figure}
\includegraphics[width=\columnwidth]{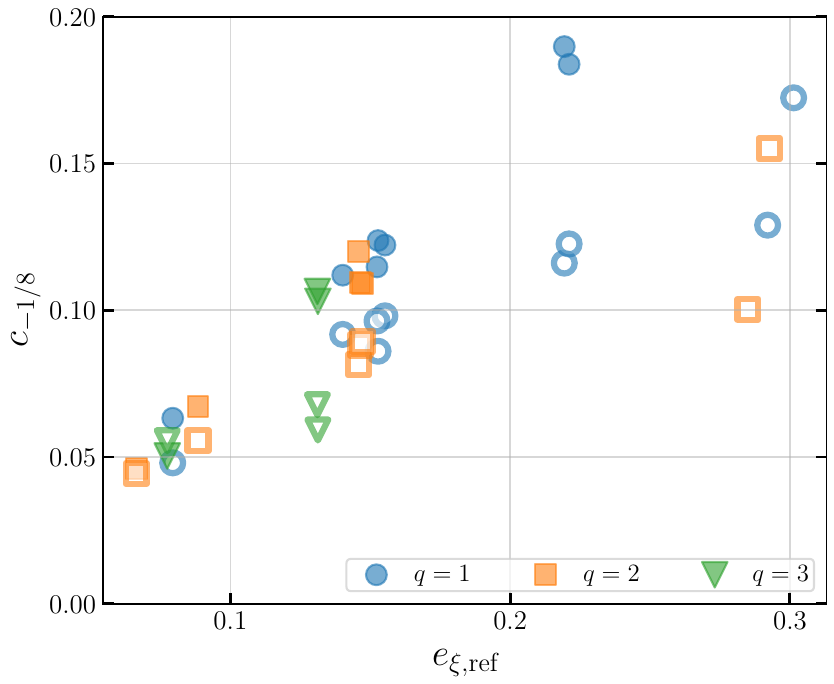}
\caption{we show the recovered best-fit coefficient $c_{-1/8}$ for the higher-order PN term with exponent $-\frac{1}{8}$ for all non-spinning SXS NR data. Filled markers represent the upper envelopes, while unfilled markers denote the lower envelopes. Details are in Section~\ref{sec:pnorders}.}
\label{fig:PNfit_param_value}
\end{figure}

To determine which higher-order terms to include, we first examine the 3PN expression in Eq.~\ref{eq:3PN_ecc}, where eccentricity is expressed as a polynomial in PN time $\tau$, and identify the missing exponents in $\tau$. This suggests that the required higher-order terms should have exponents $[-1/8, -7/8, -1]$. Each time, we then add an additional term for each exponent to the 3PN expression and fit the modulation data. Each of these fits therefore has two free parameters: initial eccentricity $e_0$ and the coefficient of the additional term. Furthermore, we explore combinations of these exponents in our fitting ansatz. We find that only the term with an exponent of $-1/8$ provides a better fit to the envelopes and remains stable near the merger. The other exponents, while fitting the envelopes, causes the fit to diverge before merger. Figure~\ref{fig:SXSBBH1371_diff_order_PNfit} illustrates this behavior, which holds for other simulations in this paper as well. Consequently, we only include the term with the exponent $-1/8$ in our final fits.

In Figure~\ref{fig:PNfit_param_value}, we show the recovered best-fit coefficient $c_{-1/8}$ for the higher-order PN term with exponent $-1/8$ for all non-spinning SXS NR data. Filled markers represent the upper envelopes, while unfilled markers denote the lower envelopes. We find that as the initial reference eccentricity $e_{\xi, \rm ref}$ increases, the best-fit $c_{-1/8}$ also increases. This indicates that, as eccentricity grows, these additional higher PN terms become more significant. This Figure also suggests that, even for $q=1$ in the low eccentricity limit, PN expressions beyond 3PN are necessary (as the coefficients to the higher PN order term is not zero). We hope this result encourages further efforts to analytically calculate the higher-order terms for eccentricity evolution within the PN framework.

\subsection{Relation between $e_{\omega_{22}}$, $e_{\rm gw}$ and ${\xi}(t)$ fits}
\label{sec:exi_to_egw}
In Section~\ref{sec:results}, we have observed that the current estimation of other eccentricity parameters, $e_{\omega_{22}}$ and $e_{\rm gw}$, using the spline-based method, is prone to unphysical behavior and fails to track well through merger (Figures~\ref{fig:gwModels_vs_gweccentricity},\ref{fig:gwModels_AlignedSpin_eccentricity}).In contrast, the estimation of \( e_{\xi} \) is consistently smooth due to its reliance on analytical fits. This raises the question of whether fits to the \( \xi(t) \) parameter could also be used to construct well-behaved versions of \( e_{\omega_{22}} \) and \( e_{\rm gw} \). Here, we first identify that $e_{\omega_{22}}$ and $e_{\rm gw}$ are related to ${\xi}$ through simple expressions. For instance, the frequencies at periastron $\omega_{22}^{\mathrm{p}}(t)$ and apastron $\omega_{22}^{\mathrm{a}}(t)$ are directly linked to the upper and lower envelopes of the modulation function ${\xi}(t)$ as:
\begin{equation}
\omega_{22}^{\mathrm{p}}(t) = \omega(t; \boldsymbol{\lambda^0}) \left(1 + \frac{\xi_{\rm p}^{\rm env}(t)}{b_{22}^{\omega}} \right),
\end{equation}
\begin{equation}
\omega_{22}^{\mathrm{a}}(t) = \omega(t; \boldsymbol{\lambda^0}) \left(1 - \frac{\xi_{\rm p}^{\rm env}(t)}{b_{22}^{\omega}}\right).
\end{equation}
The negative sign for $\xi_{\rm a}^{\rm env}(t)$ is required since we originally take its absolute value for fitting. Substituting these expressions into Eq.~(\ref{eq:ecc_omega_22}) gives:
\begin{equation}
e_{\omega_{22}}(t) = \frac{\sqrt{b_{22}^{\omega} + \xi_{\rm p}^{\rm env}(t)} - \sqrt{b_{22}^{\omega} - \xi_{\rm a}^{\rm env}(t)}}{\sqrt{b_{22}^{\omega} + \xi_{\rm p}^{\rm env}(t)} + \sqrt{b_{22}^{\omega} - \xi_{\rm a}^{\rm env}(t)}}.
\end{equation}
With some algebra, we find:
\begin{equation} 
\frac{1 - e_{\omega_{22}}^2} {2 e_{\omega_{22}}} = \frac{2 \sqrt{b_{22}^{\omega} + \xi_{\rm p}^{\rm env}(t)}\sqrt{b_{22}^{\omega} - \xi_{\rm a}^{\rm env}(t)}}{\xi_{\rm p}^{\rm env}(t) + \xi_{\rm a}^{\rm env}(t)}. 
\end{equation}
We can then use Eq.~(\ref{eq:ecc_gw}) to immediately obtain $e_{\rm gw}(t)$.

\begin{figure}
\includegraphics[width=\columnwidth]{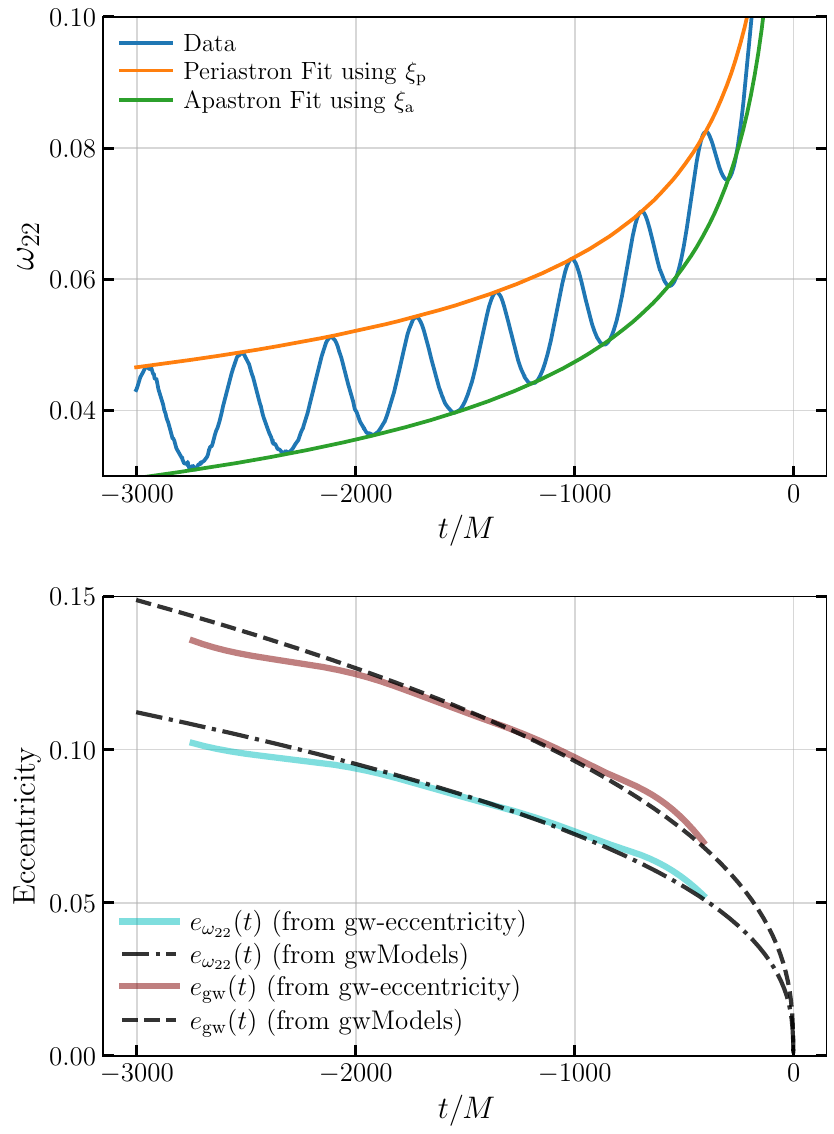}
\caption{\textit{Upper panel: } We show the $(2,2)$ mode frequency as well as the fits for the upper and lower envelopes obtained from the modulation function $\xi(t)$ for a non-spinning eccentric BBH simulation \texttt{SXS:BBH:1368}. \textit{Lower panel: }We show the estimated $e_{\omega_{22}}(t)$ and $e_{\rm gw}(t)$ obtained from \texttt{gw-eccentricity} (which uses splines) and from \texttt{gwModels} (using fits for $\xi(t)$). Details are in Section~\ref{sec:updated_egwfit}.}
\label{fig:egw_fit_better}
\end{figure}

\begin{figure}
\includegraphics[width=\columnwidth]{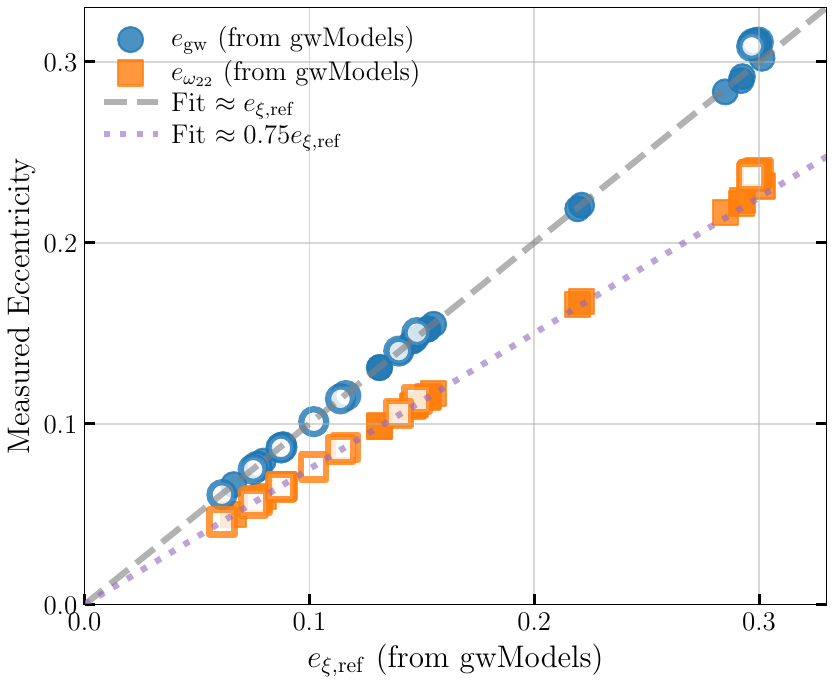}
\caption{We show the relation between $e_{\xi, \rm ref}$ with $e_{\rm gw, \rm ref}$  and $e_{\omega_{22}, \rm ref}$, after employing a robust eccentricity estimation based on $\xi(t)$ fit, for a list of SXS (solid markers) and RIT (transparent markers) non-spinning eccentric NR simulations. Details are in Section~\ref{sec:updated_egwfit}.}
\label{fig:exi_vs_egw_eomega22_nr}
\end{figure}

\subsection{Robust estimation of $e_{\omega_{22}}$ and $e_{\rm gw}$ using ${\xi(t)}$ fit}
\label{sec:updated_egwfit}
Based on the relations provided in Section~\ref{sec:exi_to_egw}, we outline the following procedure to obtain a robust, smooth fit for $e_{\omega_{22}(t)}$ and $e_{\rm gw}(t)$. First, for any eccentric waveform, we obtain the modulation function $\xi(t)$ and fit its envelopes $\xi_{\rm a}^{\rm env}(t)$ and $\xi_{\rm p}^{\rm env}(t)$ using PN-inspired expresions following the framework given in Section~\ref{sec:method}. From these, we compute $\omega_{22}^{\mathrm{p}}(t)$ and $\omega_{22}^{\mathrm{a}}(t)$ (following Section~\ref{sec:exi_to_egw}), which then yield $e_{\omega_{22}}(t)$ and $e_{\rm gw}(t)$. In Figure~\ref{fig:egw_fit_better}, we demonstrate that the expressions for $\omega_{22}^{\mathrm{p}}(t)$ and $\omega_{22}^{\mathrm{a}}(t)$ derived from the envelopes of $\xi(t)$ fit the data well for \texttt{SXS:BBH:1368} (one of the non-spinning eccentric SXS simulations considered in Section~\ref{sec:sxsnospin}). These fits are subsequently used to compute smooth $e_{\omega_{22}}(t)$ and $e_{\rm gw}(t)$, as shown in the lower panel of Figure~\ref{fig:egw_fit_better}. The updated estimates of $e_{\omega_{22}}(t)$ and $e_{\rm gw}(t)$ from \texttt{gwModels} qualitatively match the non-monotonic estimates obtained from \texttt{gw-eccentricity} and extends up to merger.

We obtain the updated and smooth $e_{\omega_{22}}$ and $e_{\rm gw}$ from \texttt{gwModels} for all non-spinning SXS simulations and a list of RIT NR data considered in Section~\ref{sec:results}. For each simulation, we note the initial reference eccentricity $e_{\xi, \rm ref}$ along with the initial eccentricities $e_{\rm gw, \rm ref}$ and $e_{\omega_{22}, \rm ref}$. In Figure~\ref{fig:exi_vs_egw_eomega22_nr}, we demonstrate that $e_{\xi, \rm ref}$ closely matches $e_{\rm gw, \rm ref}$ when employing a robust eccentricity estimation. However, their values are not identical and differs by $10^{-3}$ to $10^{-5}$. Furthermore, both $e_{\xi, \rm ref}$ and $e_{\rm gw, \rm ref}$ are larger than $e_{\omega_{22}, \rm ref}$. For the rest of the paper, we will use the framework prescribed in Section~\ref{sec:exi_to_egw} to compute $e_{\omega_{22}}$ and $e_{\rm gw}$.

\subsection{Universality in eccentricity evolution}
\label{sec:ecc_evolve}
While we can now estimate eccentricities $e_{\omega_{22}}(t)$, $e_{\rm gw}(t)$, and $e_{\xi}(t)$ from any non-precessing eccentric waveform and its corresponding circular expectation, there is currently no model for eccentricity evolution in the absence of a waveform. Astrophysical population analyses typically assume either Newtonian eccentricity evolution or fits to 1PN/2PN expectations~\cite{Vijaykumar:2024piy,Wen:2002km,Hamers:2021eir}. As a simple remedy to this problem, we first study the phenomenology of eccentricity evolution in non-spinning binaries and then present an approximate model for eccentricity evolution in Section~\ref{sec:gwEccEvNS}.

\begin{figure}
\includegraphics[width=\columnwidth]{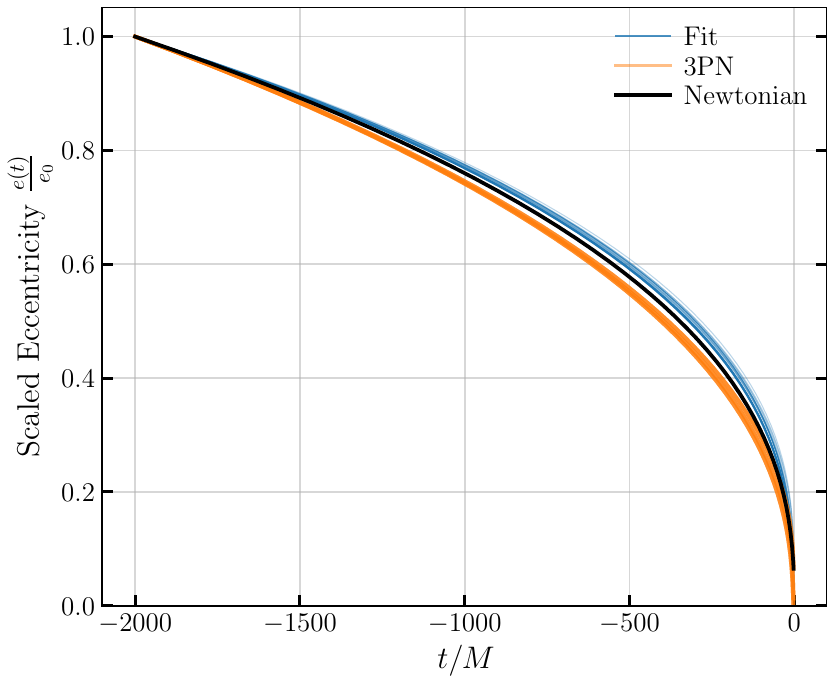}
\caption{We show the evolution of the scaled eccentricity $\frac{e(t)}{e_0}$ (blue lines), where $e_0$ is the eccentricity at the start of the waveform, for all 19 non-spinning SXS NR simulations (considered in Section~\ref{sec:sxsnospin}) over a common time window. For comparison, we also show the corresponding 3PN (orange lines) and Newtonian (black line) behaviors. Details are in Section~\ref{sec:ecc_evolve}.}
\label{fig:scaled_eccentricity_estimator}
\end{figure}

First, we note that in the Newtonian limit (given in Eq.(\ref{eq:Newt_ecc})), the evolution of eccentricity does not depend on the mass ratio. The mass ratio dependence in eccentricity evolution emerges at a higher PN order (see Eq.(\ref{eq:3PN_ecc})). Inspecting these two equations suggests that eccentricity evolution over time is either independent of or only weakly dependent on initial eccentricity values. We demonstrate this in Figure~\ref{fig:scaled_eccentricity_estimator}, where we show the scaled eccentricity $e(t)/e_{0}$, with $e(t) = e_{\xi}(t)$ and $e_0 = e_{\xi,\rm ref}$ being the reference eccentricity at the start of the waveform. The scaled eccentricity is plotted for all 19 non-spinning SXS NR simulations (used in Section~\ref{sec:sxsnospin}) over a common time window. Due to varying mass ratios ($1 \leq q \leq 3$), we observe a small spread in the estimated eccentricity evolution (blue lines). Otherwise, scaled eccentricity evolution is quite same for all simulations. For each simulation, we then input the initial eccentricity into the 3PN eccentricity evolution given in Eq.(\ref{eq:3PN_ecc}). We show these expected 3PN evolutions as orange lines. We observe similar universal evolution with a little spread. The corresponding Newtonian evolution is depicted in black. We find that the Newtonian expectation lies in between the numerical fits (blue lines) and expected 3PN evolutions (orange lines). This figure suggests that the estimated eccentricity weakly depends on mass ratio and therefore the Newtonian expression given in Eq.(\ref{eq:Newt_ecc}) could be slightly modified to capture the effective evolution of eccentricity at leading order.

\subsection{Approximate analytical model for eccentricity evolution}
\label{sec:gwEccEvNS}
\begin{figure}
\includegraphics[width=\columnwidth]{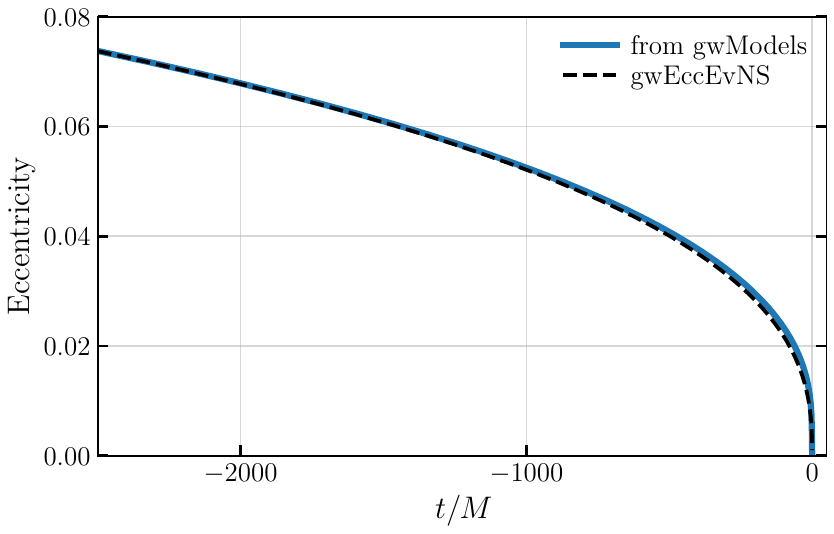}
\caption{We show the eccentricity evolution $e_{\xi}(t)$ (solid blue line; computed using the \texttt{gwModels} (\href{https://github.com/tousifislam/gwModels}{https://github.com/tousifislam/gwModels})) of the non-spinning eccentric simulation \texttt{SXS:BBH:1371} and the corresponding \texttt{gwEccEvNS} prediction (dashed black line). Details are in Section~\ref{sec:gwEccEvNS}.}
\label{fig:SXS1371_gwEccEvNS}
\end{figure}

Motivated by the near-universal behavior of eccentricity evolution discussed in Section~\ref{sec:ecc_evolve}, we develop an approximate model, \texttt{gwEccEvNS}, for the evolution of eccentricity based on our earlier fits to $e_{\xi}(t)$ from non-spinning SXS NR simulations (see Section~\ref{sec:sxsnospin}). The model is inspired by the Newtonian expression in Eq.~(\ref{eq:Newt_ecc}) and reads:
\begin{equation}
e_{\rm gwEccEvNS}(\tau, \tau_0, q, e_{0}) = e_{0} \times \left(\frac{\tau}{\tau_0}\right)^{n(q,e_0) / 48},
\label{eq:gwEccEvNS}
\end{equation}
where $n(q,e_0)$ is the numerator of the effective exponent and depends on the mass ratio $q$ and the initial eccentricity $e_0$. The dependencies are determined through fits and are given by:
\begin{align}
&n(q, e_0) =  n_1(q) n_2(e_0), \nonumber\\
&n_1(q) = -0.37857487 q + 18.4726538, \nonumber\\
&n_2(e_0) = 1 + 0.15346999 e_0 - 1.38867977 e_0^2 + 2.45635187 e_0^3.
\end{align}
We find that \texttt{gwEccEvNS} can approximately match the eccentricity evolution with an error of $\sim 5\%$. This level of agreement is sufficient for most astrophysical population studies~\cite{Vijaykumar:2024piy,Wen:2002km,Hamers:2021eir} and provides reasonable direction for waveform modeling. In Figure~\ref{fig:SXS1371_gwEccEvNS}, we show the estimated eccentricity from \texttt{gwModels} alongside the corresponding predictions from \texttt{gwEccEvNS} for \texttt{SXS:BBH:1371}. We observe similar agreement across other simulations as well. It is however possible to improve this model by considering more suitable analytical ansatz and utilizing additional NR data as they become available. We leave this for future.

\subsection{Relation between model eccentricities in PN/EOB and $\{e_{\xi}, e_{\rm gw}\}$}
\label{sec:pn_eob}
Next, we apply our framework to waveforms obtained from the PN approximation and EOB formalism. We choose the \texttt{EccentricIMR} model (used in our earlier analysis in Section~\ref{sec:sxsnospin}) within the PN framework and the \texttt{TEOBResumS} model~\cite{Chiaramello:2020ehz}~\footnote{We obtain the waveform model from the \texttt{eccentric} branch of \href{https://bitbucket.org/teobresums/teobresums}{https://bitbucket.org/teobresums/teobresums/} package} within the EOB framework. 
Similar to the NR case, we find that our method efficiently characterizes eccentricity in these waveforms as well. We confirm that, for each case, we obtain (i) smooth evolution of eccentricity and (ii) meaningful estimation of eccentricity close to merger.

\begin{figure}
\includegraphics[width=\columnwidth]{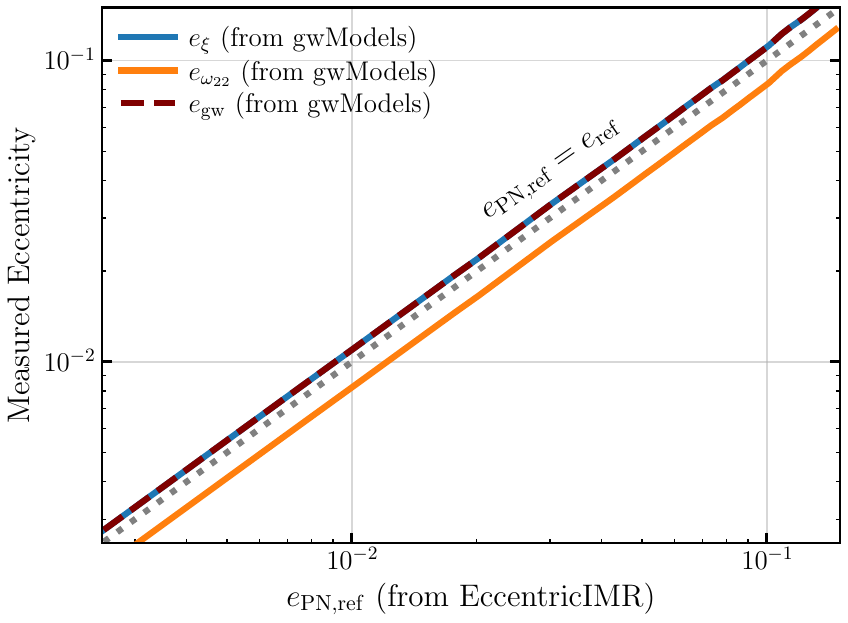}
\caption{We show the relationship between the PN reference eccentricities \( e_{\rm PN, ref} \), our eccentricity estimates $\{ e_{\xi, \rm ref}, e_{\rm gw} \}$ (computed using \texttt{gwModels}) for a total of 41 non-spinning \texttt{EccentricIMR} waveform data  with \( q = 1 \). Details are in Section~\ref{sec:pn_eob}.}
\label{fig:PNecc_vs_ecc_q1}
\end{figure}

Additionally, we investigate the relationship between the PN eccentricity in \texttt{EccentricIMR} model and the estimated waveform-based eccentricity estimators \( e_{\xi} \), \( e_{\omega_{22}} \) and \( e_{\rm gw} \). We generate 41 eccentric (and corresponding circular) waveforms for \( q = 1 \) with initial eccentricities \( 0.002 \leq e_{\rm PN, ref} \leq 0.15 \) measured at \( x_{\rm ref} = 0.07 \). Applying our eccentricity estimation framework, we obtain eccentricities \( e_{\xi} \), \( e_{\omega_{22}} \) and \( e_{\rm gw} \) at the reference frequency \( x_{\rm ref} \) using the frameworks mentioned in Section~\ref{sec:method} and ~\ref{sec:exi_to_egw}. These estimates are shown in Fig.~\ref{fig:PNecc_vs_ecc_q1}. 
We find that our eccentricity estimates \( e_{\xi, \rm ref} \) and \( e_{\rm gw, \rm ref} \) closely match each other. However, their values are (on an average 1.1 times) larger than the PN reference eccentricities. We perform simple analytical fits to the estimated reference eccentricities using \texttt{scipy.curve\_fit} module and find the following relation:
\begin{equation}
    e_{\xi,\rm ref} \approx e_{\rm gw,\rm ref} \approx 1.12 \times e_{\rm PN,ref},
\end{equation}
and
\begin{equation}
    e_{\omega_{22},\rm ref} \approx 0.85 \times e_{\rm PN,ref}.
\end{equation}
Furthermore, both \( e_{\xi, \rm ref} \) and \( e_{\rm gw, ref} \) vary smoothly with PN eccentricity. This suggests that while both definitions recover the Newtonian limit, they do not necessarily agree with PN eccentricities across the entire eccentricity range. We confirm that these observations remain the same for other mass ratio values too.

\begin{figure}
\includegraphics[width=\columnwidth]{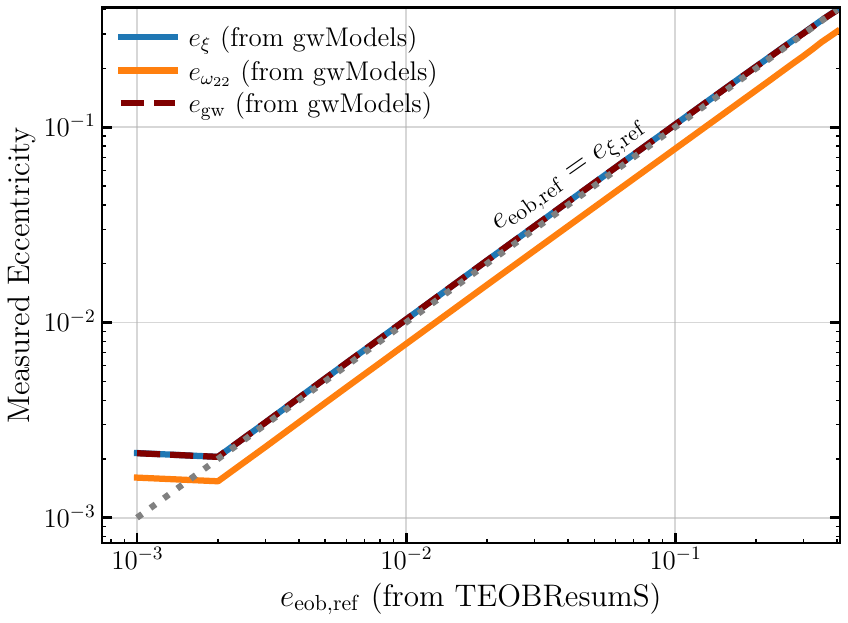}
\caption{We show the relationship between the initial eccentricities from the \texttt{TEOBResumS} model and our eccentricity estimates $\{ e_{\xi, \rm ref}, e_{\rm gw} \}$ (computed using \texttt{gwModels}) for a total of 31 non-spinning \texttt{TEOBRessumS} waveform data  with \( q = 1 \).. Details are in Section~\ref{sec:pn_eob}.}
\label{fig:eobecc_vs_eccxi}
\end{figure}

We then repeat the same exercise with \texttt{TEOBResumS} waveforms. We generate eccentric waveforms with varying initial eccentricities ($0.001 \leq e_{\rm eob, ref} \leq 0.4$) and mass ratios ($1 \leq q \leq 3$). For simplicity, we only focus on non-spinning binaries. The longest waveform in this series spans a duration of approximately $90000M$, while the shortest waveform lasts only $5000M$.  Figure~\ref{fig:eobecc_vs_eccxi} shows that our estimates for $\{ e_{\xi, \rm ref}, e_{\rm gw} \}$ match EOB eccentricities very closely. Like before, \( e_{\omega_{22, \rm ref}} \) estimates are systematically smaller than the EOB eccentricity. Additionally, we observe no noticeable mass ratio dependence on these relationship. We perform a simple analytical fit to the estimated initial eccentricities using \texttt{scipy.curve\_fit} module and find the following linear relation:
\begin{equation}
    e_{\xi,\rm ref} \approx e_{\rm gw,\rm ref} \approx e_{\rm eob,ref},
\end{equation}
and
\begin{equation}
    e_{\omega_{22},\rm ref} \approx 0.775 \times e_{\rm eob,ref}.
\end{equation}

It is important to note that, similar to the \texttt{EccentricIMR} model, the \texttt{TEOBResumS} model also relies on a 2PN radiation reaction. 
Consequently, both \texttt{EccentricIMR} and \texttt{TEOBResumS} waveforms introduce inaccuracies as they evolve over time~\cite{Hinder:2008kv,Hinder:2017sxy}. However, \texttt{TEOBResumS} includes additional calibration to NR data, which is absent in the \texttt{EccentricIMR} model. Thus, it is not surprising that the eccentricities in these two models show slightly different relationships with $e_{\rm gw}$ and $e_{\xi}$. These inaccuracies can, in principle, contribute to the differences in reference eccentricities observed in Figures~\ref{fig:PNecc_vs_ecc_q1} and~\ref{fig:eobecc_vs_eccxi}.

\begin{figure}
\includegraphics[width=\columnwidth]{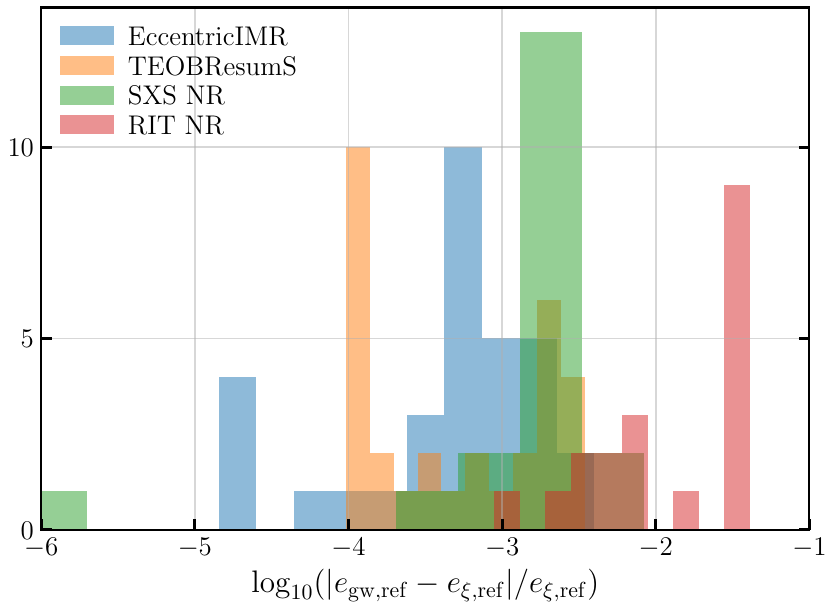}
\caption{We show the logarithm of the absolute difference between $e_{\xi, \rm ref}$ and $e_{\rm gw}$ for all \texttt{TEOBResumS} and \texttt{EccentricIMR} waveforms considered in Section~\ref{sec:pn_eob}. Additionally, we show the same for all non-spinning SXS and RIT waveforms considered in Section~\ref{sec:results}. Details are in Section~\ref{sec:pn_eob}.}
\label{fig:egw_exi_diff}
\end{figure}

Our results in Figure~\ref{fig:exi_vs_egw_eomega22_nr}, Figure~\ref{fig:PNecc_vs_ecc_q1}, and Figure~\ref{fig:eobecc_vs_eccxi} indicate that \( e_{\xi, \rm ref} \) and \( e_{\rm gw} \) closely align. To quantify this, we compute the base-10 logarithm of their relative differences for all \texttt{TEOBResumS} and \texttt{EccentricIMR} waveforms considered in this Section, as well as for all non-spinning SXS and RIT waveforms used in Section~\ref{sec:results}. These differences, shown in Figure~\ref{fig:egw_exi_diff}, are typically less than \( 10^{-2} \) and become larger than \( 10^{-2} \) for some highly eccentric RIT simulations. This consistency suggests that, despite being defined differently, both estimators yield effectively similar values for practical purposes. However, their smoothness and ability to provide eccentricity values around merger differs.

\section{Concluding remarks}
\label{sec:discussion}
In this paper, we presented a simple but robust method for characterizing gravitational waveforms from eccentric BBH mergers using only waveform quantities. Specifically, we used the universal modulation~\cite{Islam:2024rhm,Islam:2024zqo,Islam:2024bza} induced by eccentricity to build an eccentricity estimator $e_{\xi}(t)$. Using a set of non-precessing NR simulations from the SXS and RIT catalogs, as well as waveforms generated through the PN approximation and EOB formalism, we demonstrated that our framework is robust and significantly improves upon existing methods. A key feature of our method is that it is analytical and based on post-Newtonian calculations~\cite{Moore:2016qxz} with input from Newtonian expectations~\cite{Healy:2017zqj,Mroue:2010re,Maggiore:2007ulw,Purrer:2012wy}). This ensures a smooth eccentricity evolution estimation. Our framework provides not only \( e_{\xi}(t) \) but also a robust and straightforward calculation of related eccentricity estimators such as \( e_{\rm gw} \) and \( e_{\omega_{22}} \). Furthermore, it enables reliable eccentricity characterization close to merger, offering a significant improvement over existing methods.

However, the analytical calculations for eccentricity evolution are currently limited to non-spinning eccentric binaries. Despite this, our framework appears to provide smooth eccentricity measurements for aligned-spin binaries, but the effects of spin~\cite{Paul:2022xfy,Henry:2023tka} must be included in future work. 

Furthermore, the validity of the universal eccentric modulations has only been demonstrated for non-precessing binaries~\cite{Islam:2024rhm,Islam:2024zqo,Islam:2024bza}. It is essential to investigate whether precessing eccentric binaries exhibit similar relations and to determine if the eccentricity definitions presented here extend to precessing spin binaries. 

It is also important to point out that our framework requires a corresponding quasi-circular waveform for each eccentric simulation to characterize eccentricity. Different choices for the quasi-circular reference can lead to slight variations in the eccentricity estimates. In principle, it is possible to replace this requirement with a secular fit, which would make the framework more self-consistent.

Finally, we note that the effect of eccentricity should be characterized by two parameters: the eccentricity magnitude, which we have focused on in this paper, and an associated phase parameter, often represented by the mean anomaly. One can reliably extract the mean anomaly from the phase of the eccentric modulation function. We leave these investigations for the future.

Despite these limitations, our eccentricity measures are smoothly varying functions in time as well as in the binary parameter space defined by masses and spins. This makes it suitable for use in both eccentric waveform modeling and data analysis. Moreover, as this measure of eccentricity is directly related to the universal modulation function, observed in all modes and linked to eccentricity, it will simplify the process of waveform modeling even further. 

We make our framework publicly available through \texttt{gwModels} package hosted at \href{https://github.com/tousifislam/gwModels}{https://github.com/tousifislam/gwModels}.

\begin{acknowledgments}
We are grateful to the SXS collaboration and RIT NR group for maintaining publicly available catalog of NR simulations which has been used in this study. We thank Saul Teukolsky, Scott Field, Gaurav Khanna, Ajit Mehta, Vijay Varma, Keefe Mitman and Arif Shaikh for helpful discussions.
This research was supported in part by the National Science Foundation under Grant No. NSF PHY-2309135 and the Simons Foundation (216179, LB). 
Use was made of computational facilities purchased with funds from the National Science Foundation (CNS-1725797) and administered by the Center for Scientific Computing (CSC). The CSC is supported by the California NanoSystems Institute and the Materials Research Science and Engineering Center (MRSEC; NSF DMR 2308708) at UC Santa Barbara.
\end{acknowledgments}

\bibliography{References}
\newpage
\appendix
\section{3PN Eccentricity evolution}
\label{sec:3pn}
The full expression of $g_{\rm 3PN}(\tau, \tau_0, \nu)$ (used in Eq.(\ref{eq:fit_func})) up to 3PN order is~\cite{Moore:2016qxz}:
\begin{widetext}
\begin{equation}
\begin{aligned}
& g_{\rm 3PN}(\tau, \tau_0, \nu)=\left(\frac{\tau}{\tau_0}\right)^{19 / 48}\left\{1+\left(-\frac{4445}{6912}+\frac{185}{576} \nu\right)\left(\tau^{-1 / 4}-\tau_0^{-1 / 4}\right)-\frac{61}{5760} \pi\left(\tau^{-3 / 8}-\tau_0^{-3 / 8}\right)+\left(\frac{854531845}{4682022912}\right.\right. \\
&-\left.\frac{15215083}{27869184} \nu+\frac{72733}{663552} \nu^2\right) \tau^{-1 / 2}+\left(\frac{1081754605}{4682022912}+\frac{3702533}{27869184} \nu-\frac{4283}{663552} \nu^2\right) \tau_0^{-1 / 2}+\left(-\frac{19758025}{47775744}+\frac{822325}{1990656} \nu\right. \\
&\left.\quad-\frac{34225}{331776} \nu^2\right) \tau^{-1 / 4} \tau_0^{-1 / 4}+\left(\frac{104976437}{278691840}-\frac{4848113}{23224320} \nu\right) \pi \tau^{-5 / 8}+\left(-\frac{101180407}{278691840}+\frac{4690123}{23224320} \nu\right) \pi \tau_0^{-5 / 8} \\
&+\pi\left(-\frac{54229}{7962624}+\frac{2257}{663552} \nu\right)\left(\tau^{-1 / 4} \tau_0^{-3 / 8}+\tau^{-3 / 8} \tau_0^{-1 / 4}\right)+\left(-\frac{686914174175}{4623163195392}-\frac{10094675555}{898948399104} \nu+\frac{501067585}{10701766656} \nu^2\right. \\
&-\left.\frac{792355}{382205952} \nu^3\right) \tau^{-1 / 4} \tau_0^{-1 / 2}-\frac{3721}{33177600} \pi^2 \tau^{-3 / 8} \tau_0^{-3 / 8}+\left(\frac{542627721575}{4623163195392}-\frac{122769222935}{299649466368} \nu+\frac{2630889335}{10701766656} \nu^2\right. \\
&\left.-\frac{13455605}{382205952} \nu^3\right) \tau^{-1 / 2} \tau_0^{-1 / 4}+\left[\frac{255918223951763603}{186891372173721600}-\frac{15943}{80640} \gamma_E-\frac{7926071}{66355200} \pi^2+\left(-\frac{81120341684927}{13484225986560}\right.\right. \\
&\left.\left.+\frac{12751}{49152} \pi^2\right) \nu-\frac{3929671247}{32105299968} \nu^2+\frac{25957133}{1146617856} \nu^3-\frac{8453}{15120} \ln 2+\frac{26001}{71680} \ln 3+\frac{15943}{645120} \ln \tau\right] \tau^{-3 / 4} \\
&+ {\left[-\frac{250085444105408603}{186891372173721600}+\frac{15943}{80640} \gamma_E+\frac{7933513}{66355200} \pi^2+\left(\frac{86796376850327}{13484225986560}-\frac{12751}{49152} \pi^2\right) \nu-\frac{5466199513}{32105299968} \nu^2\right.} \\
&\left.\left.+\frac{16786747}{1146617856} \nu^3+\frac{8453}{15120} \ln 2-\frac{26001}{71680} \ln 3-\frac{15943}{645120} \ln \tau_0\right] \tau_0^{-3 / 4}\right\}
\end{aligned}   
\end{equation}
\end{widetext}

\section{Relationship between modulation eccentricities \{$e_{\xi}$, $e_{\xi}^{p}$, $e_{\xi}^{a}$\} and Quasi-Keplerian eccentricity $e_{t}$}
\label{sec:qk}
Based on the calculations in Refs.~\cite{Hinder:2008kv, Konigsdorffer:2006zt, Boetzel:2017zza}, Ref.~\cite{Ramos-Buades:2022lgf} derives the expression for the $(2,2)$ mode frequency $w_{22}(t)$ as a function of the post-Newtonian eccentricity $e_t$~\cite{Blanchet:2013haa} within the quasi-Keplerian framework. For the full expression, please refer to the Appendix of Ref.~\cite{Ramos-Buades:2022lgf}. The expression simplifies at the periastron and apastron points:
\begin{widetext}
\begin{equation}
\begin{aligned}
\omega_{22}^{\mathrm{a,p}}\left(x, e_t, \eta\right)=\frac{4 x^{3 / 2} \sqrt{1-e_t^2}}{\left(e_t \pm 1\right)^2\left(2 \mp e_t\right)} \pm \gamma \frac{x^{5 / 2} e_t\left(11(6 \eta-23) e_t^2 \pm(607-78 \eta) e_t+96 \eta-690\right)}{21 \sqrt{1-e_t^2}\left(e_t^2+e_t-2\right)^2} 
\end{aligned}   
\end{equation}
\end{widetext}
The upper and lower signs correspond to apastron and periastron, respectively. Substituting the expression for the $(2,2)$ mode frequencies at periastron into our eccentricity definition in Eq.~(\ref{eq:ecc_estimate}), we obtain the expression for $e_{\xi}^{\rm p}(t)$. At the lowest order, this gives:
\begin{equation}
e_{\xi}^{\rm a}(t) \approx b \left[ \frac{2 \sqrt{1-e_t^2}}{\left(e_t + 1\right)^2\left(2 - e_t\right)} - 1 \right].
\end{equation}
Expanding this in orders of $e_t$, we find:
\begin{equation}
e_{\xi}^{\rm a}(t) \approx b \left[ -e_t + \frac{23e_t^2}{8} + \cdots \right].
\end{equation}
At low eccentricity, this simplifies to $e_{\xi}^{\rm a}(t) \approx be_t$. Similarly, for low eccentricity at Newtonian order, we find $e_{\xi}^{\rm p}(t) \approx 2be_t$. Thus, we obtain:
\begin{equation}
e_{\xi}(t) \approx \frac{e_{\xi}^{\rm a}(t) + e_{\xi}^{\rm p}(t)}{2} \approx \frac{3be_t(t)}{2}.
\end{equation}
Within the post-Keplerian formalism, one can therefore have a straightforward transformation to take $e_{\xi}(t)$ to the correct Newtonian order in the low eccentricity limit as:
\begin{equation}
e_t(t) = \frac{2}{3} \frac{1}{b} e_{\xi}(t).
\end{equation}
Since we are using \( b = \frac{2}{3} \), the framework naturally reduces to the Newtonian limit at low eccentricity. However, at large eccentricities and in systems where strong-field effects are more significant, the modulation eccentricities will differ from \( e_t \). Furthermore, choosing a different value of $b$ will only linearly change the results.

\section{\texttt{gwModels} code implementation}
\label{sec:code}
Below is a code snippet demonstrating how to use \texttt{gwModels} to estimate the eccentricity from an eccentric BBH merger waveform. The package is available at \href{https://github.com/tousifislam/gwModels}{https://github.com/tousifislam/gwModels}. Detailed documentation is provided at \href{https://tousifislam.com/gwModels/gwModels.html}{https://tousifislam.com/gwModels/gwModels.html}.
\begin{widetext}
\begin{verbatim}
import gwModels

# Provide mass ratio
q = 1

# load eccentric NR data ( external function )
t_ecc, hecc_dict = get_eccentric_SXSNR_data(`SXS:BBH:1355')

# load circular NR data ( external function )
t_cir, hcir_dict = get_circular_SXSNR_data(`SXS:BBH:0180')

# compute eccentricity
obj = gwModels.ComputeEccentricity(t_ecc = t_ecc, # circular time
                                   h_ecc_dict = {'h_l2m2': hecc_dict['h_l2m2']}, # circular waveform
                                   t_cir = t_cir, # eccentric time
                                   h_cir_dict = {'h_l2m2': hcir_dict['h_l2m2']}, # eccentric waveform
                                   q=q, # mass ratio
                                   distance_btw_peaks=None, # to help peak finding
                                   t_ref = -2500, # reference time to estimate eccentricity
                                   fit_funcs_orders=['3PN_m1over8', '3PN_m1over8'], # fit PN ordeer
                                   ecc_prefactor=2/3, # eccentricity estimator pre-factor
                                   method='xi_amp', # whether to use amplitude/freq modulation
                                   include_zero_zero=False, # whether to include merger point
                                   tc = 0) # time at the merger

# reference eccentricity
obj.ecc_ref
\end{verbatim}
\end{widetext}

\end{document}